\documentclass[twocolumn,floatfix,superscriptaddress,a4paper,
                showpacs,showkeys,nofootinbib,reprint]{revtex4}
\usepackage{epsfig}
\usepackage{latexsym}
\usepackage{xspace}
\usepackage{hyperref}
\usepackage[latin2]{inputenc}
\usepackage{indentfirst}
\usepackage{enumerate}
\usepackage{color}

\usepackage{amsmath}
\usepackage{amssymb}
\usepackage[english]{babel}
\usepackage{url}
\newcommand{\bs}{\boldsymbol}

\newcommand{\eq}[1]{\begin{align} #1 \end{align}}

\begin{document}

\title{
Equations of state for real gases on the nuclear scale
}
\author{Volodymyr Vovchenko}
\affiliation{
Institut f\"ur Theoretische Physik,
Goethe Universit\"at Frankfurt, D-60438 Frankfurt am Main, Germany}
\affiliation{
Frankfurt Institute for Advanced Studies, Goethe Universit\"at Frankfurt,
D-60438 Frankfurt am Main, Germany}
\affiliation{
Department of Physics, Taras Shevchenko National University of Kiev, 03022 Kiev, Ukraine}

\date{\today}

\begin{abstract}
The formalism to augment the classical models of equation of state for real gases with the quantum statistical effects is presented.
It allows an arbitrary excluded volume procedure to model repulsive interactions, and an arbitrary density-dependent mean field to model attractive interactions.
Variations on the excluded volume mechanism  include van der Waals (VDW) and Carnahan-Starling models,
while the mean fields are based on VDW, Redlich-Kwong-Soave, Peng-Robinson, and Clausius equations of state.
The VDW parameters of the nucleon-nucleon interaction are fitted in each model to the properties of the ground state of nuclear matter,
and the following range of values is obtained: $a = 330 - 430$~MeV~fm$^3$ and $b = 2.5 - 4.4$~fm$^3$.
In the context of the excluded-volume approach,
the fits to the nuclear ground state
disfavor the values of the effective hard-core radius of a nucleon significantly smaller than $0.5$~fm, at least for the nuclear matter region of the phase diagram.
Modifications to the standard VDW repulsion and attraction terms allow to improve significantly the value of the nuclear incompressibility factor $K_0$, bringing it closer to empirical estimates.
The generalization to include the baryon-baryon interactions into the hadron resonance gas model is performed. 
The behavior of the baryon-related lattice QCD observables at zero chemical potential is shown to be strongly correlated to the nuclear matter properties: an improved description of the nuclear incompressibility also yields an improved description of the lattice data at $\mu = 0$.
\end{abstract}

\pacs{25.75.Gz, 25.75.Ag, 21.65.Mn}

\keywords{nuclear matter, real gases, van der Waals interactions}

\maketitle

\section{Introduction}
Real gas is a system of particles which interact repulsively at small distances 
and attractively at intermediate distances.
A first-order phase transition takes place
in such a system, which is a well-known phenomenon 
present in many
different molecular systems. Historically,
a cubic equation of state\footnote{A cubic equation of state is the one which can be written as a cubic function of molar volume $V_m$, or, equivalently, of specific volume $v = V/N$.
}
has often been used to describe the thermodynamic
equilibrium in the real gas.
The well-known example of such a model is the famous
van der Waals (VDW) equation of state~\cite{LL,GNS}
\eq{
\label{eq:vdworig}
p(T,V,N) = \frac{TN}{V-bN} - a\left( \frac{N}{V}\right)^2,
}
where $a>0$ and $b>0$ are the VDW parameters which describe, respectively, the attractive and the repulsive 
interactions between particles. The VDW equation \eqref{eq:vdworig} was formulated in 1873,
and for his work van der Waals obtained the 1910 Nobel Prize in Physics.
The first term in \eqref{eq:vdworig} describes the short-range repulsive interactions
by means of the excluded volume (EV) correction, whereby the system volume
is substituted by the available volume, i.e. $V \to V - bN$.
The parameter $b$ is the EV parameter. It
can be related to the classical hard-core radius of a particle
as $b = 16 \pi r^3 / 3$.
The second term
describes the attractive interactions in the mean-field approximation,
characterized by the attraction parameter $a$.
The VDW equation has underwent many different generalizations of its
repulsive and attractive terms. These had allowed to improve the description
of the thermodynamic equilibrium in the vicinity of the phase transition and critical point (CP) for
many different molecular systems. Some of these models are considered in the present paper,
and will be described in details below.

Recently, the VDW equation \eqref{eq:vdworig} was transformed to the
Grand Canonical Ensemble (GCE)~\cite{Vovchenko:2015xja},
and then generalized to include the effects of quantum statistics~\cite{Vovchenko:2015vxa,Redlich:2016dpb}.
These modifications have opened up the possibility to apply the VDW equation
in nuclear/hadronic physics, where the numbers of different particle species are usually
not conserved, and where the quantum statistical effects are often non-negligible.
In particular the nuclear matter -- a hypothetical infinite system of interacting
nucleons -- was rather successfully described by the VDW model~(see~\cite{Vovchenko:2015vxa}).
This kind of a nuclear matter description is rather different from the models which are based on the relativistic
mean-field theory, such as the Walecka model~\cite{Walecka:1974qa,Serot:1984ey} and its various generalizations~(see, e.g., Refs.~\cite{Dutra:2012mb,Dutra:2014qga,Li:2008gp} for an overview).
One notable difference is the absence of the effective mass concept in the VDW model.

The fit of the fermionic VDW equation to
the ground state properties of the symmetric nuclear matter at the normal nuclear density $n_0 = 0.16$~fm$^{-3}$ has yielded the following 
values of the VDW parameters: $a \simeq 329$~MeV~fm$^3$ and $b \simeq 3.42$~fm$^3$~\cite{Vovchenko:2015vxa}.
However, as will be detailed below, the VDW model yields a rather stiff equation
of state, with the nuclear incompressibility factor $K_0$ notably higher than the available empirical estimates. 
Thus, it is important to check 
whether these $a$ and $b$ values are robust,
in the sense that they really correctly characterize
the first interaction term in the virial expansion.
To clarify this question,
different classical equations of state for real gases,
consistent with the VDW equation in the low-density limit, are considered.
Augmented with the quantum statistical effects, they
provide a reasonable description of the nuclear matter properties.
A novel formalism, which allows to include the quantum statistical effects into the real gas models,
is introduced. It can be applied to a real gas model which contains an arbitrary EV mechanism as well as an arbitrary density-dependent mean field.

The simplicity of the presented approach allows
to include 
the VDW-like interactions into the
hadron resonance gas (HRG) model of the hadronic equation of state with relative ease. The first step in that direction was
already performed in Ref.~\cite{Vovchenko:2016rkn}, where the VDW interactions between all (anti)baryons
were included in the framework of the standard VDW equation. The parameters $a$ and $b$ were fixed to the ground state of nuclear matter.
A very strong effect of the VDW interactions on the temperature dependencies of various fluctuations of conserved charges at zero baryochemical potential, $\mu_B=0$, was reported.
A generalization to arbitrary forms of the EV and of the density-dependent mean field is performed in the present work.
Strong correlation between the behavior of the baryon-related HRG observables and the nuclear matter properties is demonstrated.
This suggests an intriguing possibility
to see the effects of the VDW-like interactions between hadrons in lattice QCD~\cite{Borsanyi:2011sw,Bazavov:2012jq} simulations, 
as well as in the fluctuation measurements in heavy-ion collision experiments.

The paper is organized as follows.  
Classical
equations of state of real gases,
which are considered in the present work, 
are listed in Sec.~\ref{sec-vdw}.
A quantum statistical generalization for an arbitrary real gas fluid
is elaborated in Sec.~\ref{sec-full}.
In Sec.~\ref{sec-nm} the real gas models are applied to the description of the
symmetric nuclear matter. A comparable analysis of the different real gas models is presented.
An extension of the HRG model to include the baryonic interactions in the framework of the real gas models is described in Sec.~\ref{sec-RGHRG}.
A summary in Sec.~\ref{sec-summary}
closes the article.

\section{Classical equations of state of real gases}
\label{sec-vdw}
Classical equations of state for real gases are listed in this section. These equations are written in terms of the pressure $p$ as a function of the temperature $T$ and the particle density $n$. They are subsequently used in the present work.

\subsection{van der Waals}
The first model under consideration is the aforementioned \emph{van der Waals} equation of state~\eqref{eq:vdworig}
\eq{\label{eq:vdwcl}
p(T,n) = \frac{Tn}{1-bn} - a\,n^2,
}
which is rewritten here as the pressure in terms of the temperature $T$ and the particle density $n \equiv N/V$.

Over the years, many modifications of the original VDW equation were developed.
This concerns both the attractive term and the repulsive term.
Modifications to the attractive term include the addition of an explicit dependence
on the repulsion parameter $b$, and also a
temperature dependence of the attraction parameter $a$.
The present work deals with the the nuclear matter at low temperatures, including the $T=0$ region.
For this reason the possibility of the $T$-dependent parameter $a$ is omitted in the present paper.

\subsection{Redlich-Kwong-Soave}
One historically well-known modification of the VDW equation is the
\emph{Redlich-Kwong} equation of state~\cite{RedlichKwong} with the \emph{Soave}
modification~\cite{Soave}. 
The classical equation of state for this RKS model reads
\eq{\label{eq:vdwrks}
p(T,n) = \frac{Tn}{1-bn} - \frac{a\,n^2}{1+bn}.
}
The parameter $a$ is temperature-dependent in the original formulation of the RKS equation.
As mentioned above, such a possibility is not considered in the present work.
Thus, the parameter $a$ is taken to be temperature independent.

\subsection{Peng-Robinson}
Another popular modification is the \emph{Peng-Robinson} (PR) equation of state~\cite{PengRobinson}.
The PR pressure reads
\eq{\label{eq:vdwpr}
p(T,n) = \frac{Tn}{1-bn} - \frac{a\,n^2}{1 + 2bn - (bn)^2}.
}

\subsection{Clausius}
Finally, the model based on the \emph{Clausius} equation is also considered
\eq{\label{eq:vdwclausius}
p(T,n) = \frac{Tn}{1-bn} - \frac{a\,n^2}{(1+cn)^2}.
}
In general, this is a three-parameter model.
The parameter $c$ has the unit of volume.
Unless stated otherwise, only the two-parameter models are considered in the present work, i.e. one assumes $c \equiv b$. Note also that the parameter $a$ is temperature dependent in the original Clausius model: typically it is inversely proportional to $T$. 
Similar to the previous two models, this temperature dependence is omitted in the present work.

\subsection{Carnahan-Starling modification of the repulsive term}
The first term in Eqs.~\eqref{eq:vdwcl}-\eqref{eq:vdwclausius} corresponds to the VDW excluded volume correction. 
It is clear that 
the maximum value of the particle density (packing limit) 
in such a model
is restricted from above: $bn < 1$.
The VDW repulsive term can be improved.
One simple but powerful modification is the \emph{Carnahan-Starling} (CS) model formulated in 1969~\cite{CarnahanStarling}.
The pressure in the CS model reads
\eq{\label{eq:cs}
p^{\rm cs} (T,n) = Tn \, \frac{1+\eta+\eta^2-\eta^3}{(1-\eta)^3},
}
where $\eta = bn / 4 = 4 \pi n r^3 / 3$ is the so-called \emph{packing fraction} -- the fraction of the total system volume occupied by the total cumulative volume of all finite-sized spherical particles in the system.
The relation \eqref{eq:cs} gives a very good approximation
of the pressure of the thermodynamic system of particles with the classical hard sphere
interaction. It gives an adequate description of the hard spheres equation of state up to much higher densities compared
to the standard VDW model. Recently this model was successfully used in some
hadronic physics
applications~\cite{Anchishkin:2014hfa,Satarov:2014voa,Satarov:2016peb}.

Substituting the first term in Eqs.~\eqref{eq:vdworig}-\eqref{eq:vdwclausius} with $p^{\rm cs} (T,n)$~\eqref{eq:cs} one obtains four additional real gas models: the VDW-CS, RKS-CS, PR-CS, and Clausius-CS models.


It is clear that the models with the CS repulsive term have a
packing limit of $bn<4$, which is larger than in the models with the VDW repulsive term.
The only exception is the PR-CS model, where due to the presence of a pole in the denominator of the attraction term one has a lower limit $bn < 1+\sqrt{2}$.

It is evident that all considered models reduce to the VDW equation in the low-density limit. More specifically, they all coincide up to the 2nd order virial expansion. 
All of the models also predict the existence of the liquid-gas phase transition and of the associated CP. With regards to the critical behavior in the vicinity of the CP they all fall into the universality class of the mean field theory.
However, quantitative difference between the models is non-negligible at high densities, in particular in the vicinity of the CP and/or in the liquid phase. 

\section{Quantum statistical formulation}
\label{sec-full}

\subsection{Free energy}
The classical equations of state listed in the previous section
are all formulated in the canonical ensemble (CE). They are given in terms of the pressure $p$ as a function
of the temperature $T$ and the particle density $n \equiv N/V$.
This knowledge of the equation of state, however, is insufficient
to obtain full thermodynamic information about the system.
In particular, it gives no information about the mass or the degeneracy of the particle.
The reason for this is that the $T$, $V$, and $N$ variables are not the natural variables for the pressure function.
The thermodynamical potential in the CE is the free energy $F(T,V,N)$, and for a complete thermodynamic description one has to reconstruct this quantity. 
The free energy satisfies the following
equation:
\eq{
\label{eq:Fderiv}
\left( \frac{\partial F}{\partial V} \right)_{T,N} = -p(T,V,N).
}

All models under consideration can be described
by the free energy of the following form
\eq{\label{eq:Fgen}
F_{\rm cl}(T,V,N) = F^{\rm id}_{\rm cl} (T, V f(\eta), N) + N\,u(n),
}
where $F^{\rm id}_{\rm cl}$ is the free energy of the corresponding ideal gas, and where the subscript \emph{cl} corresponds to the classical (Maxwell-Boltzmann) statistics.
As before, $\eta = bn/4$ and $n \equiv N/V$.
The function $f(\eta)$ quantifies the fraction of the total
volume which is available for particles to move in at
the given value of the packing fraction (density) $\eta$.
Evidently, it has to take the values in the range $0 < f(\eta) \leq 1$.
The quantity $u(n)$ in \eqref{eq:Fgen} is the self-consistent density-dependent
\emph{mean field}\footnote{Technically, the $u(n)$ is the mean total energy of attractive interactions per particle.}.
The $u(n)$ describes
the attractive interactions in the models under consideration. 

Taking the derivative of $F_{\rm cl}(T,V,N)$ with respect to the volume [Eq.~\eqref{eq:Fderiv}], one
obtains the following general expression for the CE pressure
\eq{\label{eq:pgencl}
p_{\rm cl}(T,n) = \frac{nT}{f(\eta)} [f(\eta) - \eta\, f'(\eta)] + n^2\,u'(n),
}
where the relation $[\partial F^{\rm id}_{\rm cl} (T,V,N) / \partial V]_{T,N} = -n\,T$ was used.

Comparing \eqref{eq:pgencl} with the equations of state listed in Sec.~\ref{sec-vdw} one can reconstruct the functions $f(\eta)$ and $u(n)$ for all models.
Depending on the EV model one obtains the following for $f(\eta)$:
\begin{enumerate}
\item van der Waals EV
$$
f(\eta) = 1 - 4\eta .
$$
\item Carnahan-Starling EV
$$
f(\eta) = \exp\left(-~\frac{(4-3\eta) \eta}{(1-\eta)^2}\right) .
$$
\end{enumerate}

The list of different possibilities for $u(n)$ is the following:
\begin{enumerate}
\item van der Waals
$$
u(n) = -a\,n .
$$
\item Redlich-Kwong-Soave
$$
u(n) = -\frac{a}{b} \,\log(1+bn) .
$$
\item Peng-Robinson
$$
u(n) = -\frac{a}{2\sqrt{2}b}\,\log\left(\frac{1+bn+\sqrt{2}bn}{1+bn-\sqrt{2}bn}\right) .
$$
\item Clausius
$$
u(n) = -\frac{an}{1+cn} .
$$
\end{enumerate}

\subsection{Quantum statistics}

The classical equations of state for real gases do not include quantum statistical effects. This is usually not a problem
for the description of the liquid-vapour equilibrium in molecular
systems for which these models were originally designed.
However, in the nuclear matter the situation is very different: the ground
state of nuclear matter is located at $T=0$ and the critical temperature is $T_c=15-20$ MeV.
In this range the effects of Fermi statistics are crucial, and the application of a classical model
will lead to unphysical results, such as the negative values of the entropy density~(see \cite{Vovchenko:2015vxa} for details). 
The VDW equation with Fermi statistics was
formulated in Ref.~\cite{Vovchenko:2015vxa}.
To my knowledge, similar formulations for other real gas models are presently missing.
Following Ref.~\cite{Vovchenko:2015vxa}, one can list some general conditions that a quantum statistical generalization of the classical VDW-like equation of state must satisfy:
\begin{enumerate}
\item It should be transformed to the ideal quantum gas
at $a=0$ and $b=0$.
\item It should be equivalent
to the corresponding classical equation of state
in a region of the thermodynamical parameter values where the quantum statistical effects can be neglected.
\item The entropy should be a non-negative quantity and go to zero at $T\rightarrow 0$.
\end{enumerate}

The following \emph{ansatz} for the free energy of the quantum statistical equation of state of a real gas is assumed: the $F(T,V,N)$ has the form given by Eq.~\eqref{eq:Fgen}, i.e.
\eq{\label{eq:Fquant}
F(T,V,N) = F^{\rm id} (T, V f(\eta), N) + N\,u(n),
}
but here the $F^{\rm id} (T, V, N)$ is the free energy of the corresponding
\emph{quantum} ideal gas. It is evident that 
the first two conditions
are
satisfied by construction. 
Note that the exact analytic form of $F^{\rm id} (T, V, N)$ is not specified here.
As will be seen below, the present calculations do not require this.
In order to check the 
third condition
one can calculate the entropy. Such a calculation yields
\eq{
S(T,V,N) = -\left(\frac{\partial F}{\partial T}\right)_{V,N} = S^{\rm id} (T, V f(\eta), N),
}
where $S^{\rm id} (T, V, N)$ is the entropy of the corresponding ideal quantum gas. The $S(T,V,N)$ is always non-negative and goes to zero at $T \to 0$. Thus,
the model satisfies the third condition.
Other quantities in the CE can be obtained by the standard thermodynamic relations. In particular one can calculate the energy
\eq{
E(T,V,N) = F + TS =  E^{\rm id} (T, V f(\eta), N) + N\,u(n),
}
the pressure
\eq{
p(T,V,N) &= -\left(\frac{\partial F}{\partial V}\right)_{T,N} \nonumber \\
&= p^{\rm id} (T, V f(\eta), N) \, [f(\eta) - \eta\, f'(\eta)] 
\nonumber \\
& \quad  +  n^2 \, u'(n),
}
and the chemical potential
\eq{
\mu(T,V,N) &= \left(\frac{\partial F}{\partial N}\right)_{T,V} 
\nonumber \\
&= \mu^{\rm id} (T, V f(\eta), N) 
\nonumber \\
& \quad
- \frac{b}{4} \, f'(\eta) \, p^{\rm id} (T, V f(\eta), N)
\nonumber \\
& \quad
+ u(n) + n\, u'(n).
}

All intensive quantities in the CE depend only on the
temperature $T$ and the particle density $n$ in the thermodynamic limit. Thus, one can write these quantities as the following
\eq{\label{eq:TDCE}
p(T,n) &= p^{\rm id}_{_{\rm CE}} \left(T, \frac{n}{f(\eta)}\right) \, [f(\eta) - \eta\, f'(\eta)] + n^2 \, u'(n), \\
s(T,n) &= \frac{S}{V} = f(\eta) \, s^{\rm id}_{_{\rm CE}} \left(T, \frac{n}{f(\eta)}\right), \\
\varepsilon(T,n) & = \frac{E}{V} = f(\eta) \, \varepsilon^{\rm id}_{_{\rm CE}} \left(T, \frac{n}{f(\eta)}\right) + n \, u(n), \\
\label{eq:TDCE-end}
\mu(T,n) &= \mu^{\rm id}_{_{\rm CE}} \left(T, \frac{n}{f(\eta)}\right) - \frac{b}{4} \, f'(\eta) \, p^{\rm id}_{_{\rm CE}} \left(T, \frac{n}{f(\eta)}\right) 
\nonumber \\
& \quad
+ u(n) + n\, u'(n),
}
where the functions with the superscript \emph{id} and the subscript \emph{CE} denote the corresponding CE functions of the ideal quantum gas.

For the particular case of the VDW equation, the presented quantum statistical formulation is mathematically equivalent to the ones previously proposed in Refs.~\cite{Vovchenko:2015vxa,Redlich:2016dpb}.
Note that the obtained CE pressure \eqref{eq:TDCE} is now different from the classical CE pressure~\eqref{eq:pgencl}
because of the quantum statistical effects. 
Equation \eqref{eq:TDCE} reduces to \eqref{eq:pgencl} whenever the Boltzmann approximation is applied.

\subsection{Grand canonical ensemble}
In the CE,
the number of particles $N$ is an independent variable, and this number is fixed exactly in each microscopic state. On the other hand, the number of different hadron species is usually not conserved in the hadronic physics applications. In this case it is often more convenient to work in the \emph{grand canonical ensemble} (GCE), where only an average number of particles is fixed. This value is regulated by the chemical potential $\mu$. 
The particle density therefore becomes a function of $T$ and $\mu$ in the GCE, i.e. $n = \langle N \rangle / V = n(T,\mu)$. The GCE pressure $p(T,\mu)$, expressed as a function of its natural variables $T$ and $\mu$, contains the complete information about the system.

Both the CE and the GCE are equivalent for calculating the averages in the thermodynamic limit. This thermodynamic equivalence of ensembles, however, does not extend to the fluctuations.
For example, the GCE formulation 
allows to calculate the particle number fluctuations in the vicinity of the CP, which by definition are absent in the CE.
The GCE formulation also allows a generalization to the multi-component hadron gas,
where the numbers of different hadronic species are not conserved.
This opens new applications in the physics of heavy-ion collisions and QCD equation of state~(see Sec.~\ref{sec-RGHRG} below).

To the best of my knowledge, the GCE formulation has been missing in the literature even for the classical (Maxwell-Boltzmann) real gas equations of state. Thus, the GCE formulation of the quantum statistical real gas models is briefly described in this subsection.


The thermodynamic equivalence of ensembles with regards to the averages will be employed to transform the models for the equation of state of real gases from the CE to the GCE.
First, the following notations are introduced:
\eq{
p^* = p^{\rm id}_{_{\rm CE}} \left( T, \frac{n}{f(\eta)} \right), 
\quad
& n^* = n^{\rm id}_{_{\rm CE}} \left( T, \frac{n}{f(\eta)} \right), 
\\
s^* = s^{\rm id}_{_{\rm CE}} \left( T, \frac{n}{f(\eta)} \right), 
\quad
& \mu^* = \mu^{\rm id}_{_{\rm CE}} \left( T, \frac{n}{f(\eta)} \right),
}
where the ideal gas functions correspond to the CE.
From the expression for the $\mu^*$ 
it follows that
\eq{\label{eq:nns}
\frac{n}{f(\eta)} = n^{\rm id} (T, \mu^*),
}
where $n^{\rm id} (T, \mu^*)$ is the GCE ideal gas density at the temperature $T$ and chemical potential $\mu^*$. 
It then follows from \eqref{eq:nns} that
\eq{
p^* = p^{\rm id} \left( T, \mu^* \right), 
\,\,
n^* = n^{\rm id} \left( T, \mu^* \right), 
\,\,
s^* = s^{\rm id} \left( T, \mu^* \right), 
}
where all the ideal gas functions now correspond to the GCE.
The corresponding ideal gas relations are listed in Appendix~\ref{app-ig}.

Using Eqs.~\eqref{eq:TDCE}-\eqref{eq:TDCE-end} one can now write all the thermodynamic quantities for the equation of state of a real gas in the GCE
\eq{\label{eq:TDGCE}
p(T,\mu) &= [f(\eta) - \eta\, f'(\eta)] \, p^{\rm id}(T,\mu^*) + n^2 \, u'(n), \\
s(T,\mu) &= f(\eta) \, s^{\rm id}(T,\mu^*), \\
\varepsilon(T,\mu) & = f(\eta) \, \varepsilon^{\rm id}(T,\mu^*) + n \, u(n), \\
\label{eq:TDGCE-end}
n(T,\mu) &= f(\eta) \, n^{\rm id}(T,\mu^*).
}
In the above equations it is always implied that $n \equiv n(T,\mu)$.
It is evident from Eqs.~\eqref{eq:TDGCE}-\eqref{eq:TDGCE-end} that once the ``shifted'' chemical potential $\mu^* = \mu^* (T,\mu)$ is known
for the given $T$ and $\mu$, all the other quantities can be determined. 
Indeed, first the $n(T,\mu)$ is calculated by numerically solving Eq.~\eqref{eq:TDGCE-end}. Then the calculation of all other quantities is straightforward.
The $\mu^* (T,\mu)$ itself is obtained as the solution to the following transcendental equation
\eq{\label{eq:mustar}
\mu  = \mu^* - \frac{b}{4} \, f'(\eta) \, p^* + u(n) + n\, u'(n).
}
This equation follows from Eq.~\eqref{eq:TDCE-end}.

The algorithm for the calculations within the GCE is the following: (i) at a given $T$-$\mu$ pair Eq.~\eqref{eq:mustar} is solved numerically to determine the $\mu^*$; (ii) all other quantities are calculated using Eqs.~\eqref{eq:TDGCE}-\eqref{eq:TDGCE-end};
(iii) if there are multiple solutions to Eq.~\eqref{eq:mustar} for a given $T$-$\mu$ pair, then the solution with the largest pressure is chosen, in accordance with the Gibbs criterion.

The possible appearance of multiple solutions to Eq.~\eqref{eq:mustar} at a given $T$-$\mu$ pair is related to the irregular behavior of the VDW-like isotherms below the critical temperature $T_c$ of the first-order phase transition. The application of the Gibbs criterion in such a case is equivalent to the Maxwell construction of equal areas in the CE~(see Ref.~\cite{Vovchenko:2015vxa} for more details).

In the particular case of the VDW equation with $a=0$ one obtains $\mu^* = \mu - b\,p$ and $p(T,\mu) = p^{\rm id} (T, \mu - b\,p)$. This reproduces the GCE excluded volume model formulated in Ref.~\cite{Rischke:1991ke}.

It should be noted that the present formalism is not restricted exclusively to the real gas models. In fact, it can be applied to any system for which the free energy can be written in the form given by Eq.~\eqref{eq:Fquant}. In particular, the density dependent mean field $u(n)$ does not necessarily have to be attractive, but can contain both attractive and repulsive interactions.
Similarly, the presence of the EV corrections is a possibility, but not a necessity of this formalism.

\section{Nuclear matter} 
\label{sec-nm}
In this section the quantum statistical models of the equation of state for real gases are applied to describe the
properties of the symmetric nuclear matter.
Namely, the Fermi gas of nucleons (with mass $m\cong 938$~MeV and (iso)spin degeneracy $d=4$) is considered. 
At the same time, the formation of the nucleon clusters (i.e., ordinary nuclei) will be neglected.
The attractive and repulsive interactions between nucleons are
described by the VDW-like parameters $a$ and $b$, respectively.
Eight different real gas models, listed in Sec.~\ref{sec-vdw}, are employed:
the VDW, RKS, PR, Clausius, VDW-CS, RKS-CS, PR-CS, and Clausius-CS models.
The description of the nuclear matter properties is an important first step. In particular, it allows to obtain the restrictions on the values of the interaction parameters $a$ and $b$. This step has to be considered before further applications to the physics of the hadron resonance gas model, the QCD equation of state, or heavy-ion collisions can be performed. 

A study of nuclear matter is certainly not a new subject.
The thermodynamics of nuclear matter and its applications to 
the production of the nuclear fragments in heavy ion collisions
were considered in Refs. \cite{Jennings:1982wi,Ropke:1982ino,Fai:1982zk,Biro:1981es,Stoecker:1981za,Csernai:1984hf} in 1980s (see Ref.~\cite{Csernai:1986qf} for a review of these early developments).
Nowadays, the properties of nuclear matter are described
by many different models,
particularly by those which employ the relativistic mean-field theory~\cite{Serot:1984ey,Zimanyi:1990np,Brockmann:1990cn,Mueller:1996pm,Bender:2003jk}.
Earlier, the EV corrections had already been considered in the mean-field models, where they were added on top of the repulsive force described by the $\omega$ meson exchange~\cite{Rischke:1991ke,Anchishkin:1995np}, or on top of the Skyrme-type density dependent repulsive mean field~\cite{Satarov:2009zx}.
In the present work, however, the repulsive forces are described \emph{solely} by the EV corrections.

Experimentally, the presence of the liquid-gas phase transition
in nuclear matter was first reported in Refs.~\cite{Finn:1982tc,Minich:1982tb,Hirsch:1984yj} 
by indirect observations. The direct measurements of the nuclear caloric 
curve were first done by the ALADIN collaboration~\cite{Pochodzalla:1995xy}, and were
later followed by other experiments~\cite{Natowitz:2002nw,Karnaukhov:2003vp}.

%

\subsection{Fits to the ground state of nuclear matter}
Each of the eight models under consideration contains two VDW-like parameters $a$ and $b$. The values of these parameters need to be fixed. In molecular systems, these parameters are usually fixed in order to reproduce the experimentally known location of the CP.
A different strategy is employed for the nuclear matter: the parameters $a$ and $b$ are fixed to reproduce the properties of nuclear matter in its ground state. At $T=0$ and $n=n_0\cong 0.16$~fm$^{-3}$ one has $p=0$ and $\varepsilon / n = m + E/A \cong 922$~MeV~(see, e.g., Ref.~\cite{Bethe:1971xm}). 
Here $E/A \cong - 16$~MeV denotes the binding energy per nucleon.
Once the parameters $a$ and $b$ are fixed, the location of the CP of the nuclear liquid-gas transition becomes a prediction of the model, which can be compared to the experimental estimates.

The following values of parameters $a$ and $b$ were obtained in Ref.~\cite{Vovchenko:2015vxa} from the fit to the nuclear ground state within the VDW model: $a \cong 329$~MeV~fm$^3$ and $b \cong 3.42$~fm$^3$. 
The values of $a$ and $b$, however, can be different for other real gas models.
While all considered real gas models are consistent with the VDW equation in the low-density limit ($n \ll n_0$), the ground state of nuclear matter which is used to fix the parameters cannot be considered as belonging to the low-density limit. 
Indeed, at $n = n_0 = 0.16$~fm$^3$ the resulting packing fraction $\eta_0 = b \, n_0 / 4 \simeq 0.14$ is larger than the half of the VDW packing limit of 0.25.
Because the pressure in the classical VDW excluded volume model rises sharply at large densities, the applicability of the model may be doubtful in that region. 
The VDW model does predict reasonable values of the critical temperature $T_c \cong 19.7$~MeV and the critical density $n_c \cong 0.072$~fm$^{-3}$.
However, the model itself appears to give a rather stiff equation of state.
This stiffness 
can be characterized by the nuclear ``incompressibility'' factor $K_0 = 9 \, ( \partial P / \partial n)_{T}$, calculated at the normal nuclear density. The VDW model gives $K_0 \cong 763$~MeV, a value significantly higher than the recent empirical estimate of $250\,-\,315$~MeV~\cite{Stone:2014wza}, or a value of about 550~MeV obtained in the standard Walecka model~\cite{Dutra:2014qga}.  Thus, it is evident that extensions of the VDW model are needed.
Application of the Carnahan-Starling EV model is particularly interesting in that regard. It is known to have a significantly higher applicability range in the classical physics compared to the VDW excluded volume model.

\begin{figure}[t]
\centering
\includegraphics[width=0.49\textwidth]{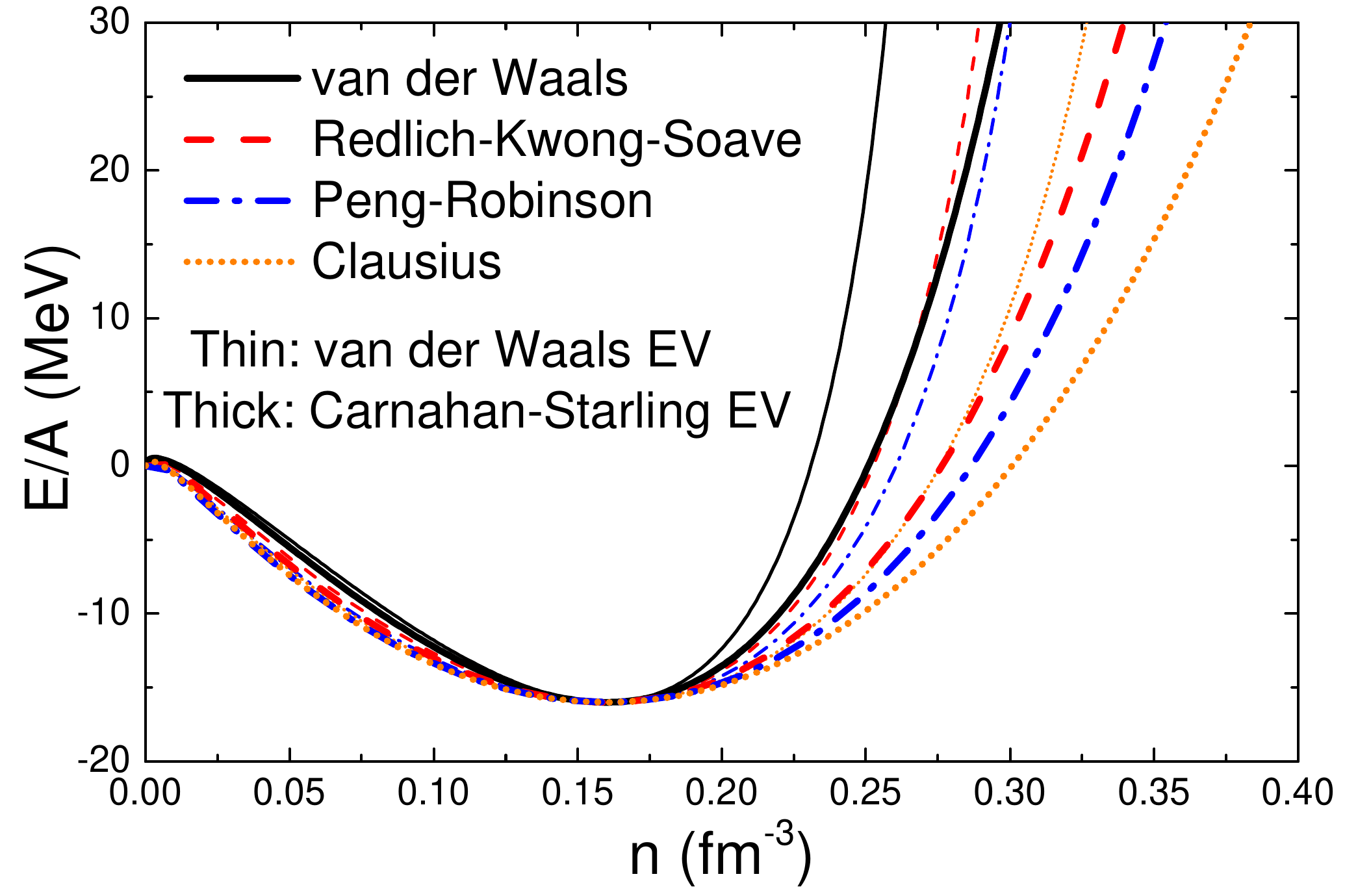}
\caption[]{
The nucleon number density dependence of the binding energy per nucleon $E/A$ in the symmetric nuclear matter calculated within eight different models of real gas at $T=0$. The thin lines denote calculations within the four models with the VDW excluded volume term, i.e. they correspond to the VDW (solid black line), RKS (dashed red line), PR (dash-dotted blue line), and Clausius (dotted orange line) models.
The thick lines correspond to the models with the Carnahan-Starling excluded volume term.
}\label{fig:EA}
\end{figure}

The values of the VDW parameters $a$ and $b$ obtained from the fit to the ground state of nuclear matter within eight different models of real gases, as well as the resulting incompressibility $K_0$, are listed in Table~\ref{tab:comp}. 
The density dependence of the binding energy per nucleon $E/A$ at zero temperature is shown in Fig.~\ref{fig:EA}.
The modifications to the attractive and repulsive VDW terms affect strongly the resulting values of the incompressibility factor: among all of the considered models, the $K_0$ is the largest in the VDW model. 
The $K_0$ attains systematically smaller values whenever the Carnahan-Starling EV model is used instead of the VDW excluded volume model. This is expected: the CS model provides a softer equation of state.
As seen in Fig.~\ref{fig:EA}, the softest equation of state is given by the Clausius-CS model. This model yields $K_0 = 333$~MeV, which is  the smallest value among all of the considered models.
This value is still larger than the recently reported empirical range of $250\,-\,315$~MeV~\cite{Stone:2014wza}, however, it is in a much better agreement compared to the original VDW model. 

The CS modification has the advantage over the VDW excluded volume model regarding the high-density behavior.
In particular, the problems with causality, typical for the standard VDW excluded volume approach at high densities, may only emerge at significantly higher densities in the CS model.
Thus, the CS modification extends the applicability range of a real gas model.
Similar conclusion was obtained in Ref.~\cite{Satarov:2014voa} for a high-temperature Boltzmann gas of hadrons.

The present formalism also allows to consider a real gas model with more than two interaction parameters.
For example, the original Clausius/Clausius-CS model contains three interaction parameters: $a$, $b$, and $c$.
By relaxing the assumption $c \equiv b$, employed in the present work,
one can improve the description of the $K_0$, and of other nuclear matter properties.
For instance, the three-parameter Clausius model yields $K_0 \simeq 315$~MeV for $a = 437$~MeV~fm$^3$, $b = 2.14$~fm$^3$, $c  \simeq 1.64 \, b = 3.51$~fm$^3$, and $K_0 \simeq 250$~MeV for $a = 472$~MeV~fm$^3$, $b = 1.73$~fm$^3$, $c  \simeq 2.74 \, b = 4.74$~fm$^3$.
Similarly, the three-parameter Clausius-CS model gives $K_0 \simeq 315$~MeV for $a = 431$~MeV~fm$^3$, $b = 2.66$~fm$^3$, $c  \simeq 1.17 \, b = 3.11$~fm$^3$, and $K_0 \simeq 250$~MeV for $a = 469$~MeV~fm$^3$, $b = 2.03$~fm$^3$, $c  \simeq 2.20 \, b = 4.47$~fm$^3$.
Real gas models with more than two interaction parameters will be considered in more details elsewhere. 


The values of the VDW-like parameters $a$ and $b$ depend on the model used to fix them. The VDW model yields $a \simeq 329$~MeV~fm$^3$.
This is the smallest value of the attraction parameter among all considered models. The Clausius-CS and PR-CS models do a better job in describing the $K_0$. These models yield $a=420-430$~MeV~fm$^3$.
This suggests that the fit of the nuclear ground state within the standard VDW model underestimates the value of $a$, at least in the context of the low-density virial expansion.

\begin{table*}
 \caption{The values of the VDW-like parameters $a$ and $b$, 
 the values of the resulting nuclear incompressibility, and the properties of the CP of nuclear matter. These quantities are listed for eight different real gas models. The corresponding empirical estimates, as well as the results of the Walecka model, are  listed as well. } 
 \centering                                                 
 \begin{tabular*}{\textwidth}{c @{\extracolsep{\fill}} |cc|c|ccccc}                                   
 \hline
 \hline
 Model & $a$~(MeV~fm$^3$) & $b$~(fm$^3$) & $K_0$~(MeV) & $T_c$~(MeV) & $n_c$~(fm$^{-3}$) & $p_c$~(MeV/fm$^3$) & $Z_c = \frac{p_c}{n_c \, T_c}$ \\
 \hline
 VDW          & 329  & 3.42 & 763 & 19.7 & 0.072 & 0.52 & 0.37 \\
 RKS          & 374  & 2.94 & 518 & 18.2 & 0.064 & 0.40 & 0.34 \\
 PR           & 408  & 2.82 & 443 & 17.1 & 0.061 & 0.33 & 0.32 \\
 Clausius     & 407  & 2.50 & 390 & 17.5 & 0.059 & 0.33 & 0.32 \\
 \hline
 VDW-CS       & 347  & 4.43 & 601 & 18.6 & 0.070 & 0.47 & 0.36 \\
 RKS-CS       & 394  & 3.50 & 421 & 17.4 & 0.061 & 0.35 & 0.33 \\
 PR-CS        & 433  & 3.37 & 359 & 16.2 & 0.057 & 0.29 & 0.31 \\
 Clausius-CS  & 423  & 2.80 & 333 & 16.9 & 0.056 & 0.30 & 0.32 \\
 \hline
 Walecka~\cite{Silva:2008zza}	  
 			  & --   & --   & 551 & 18.3 & 0.065 & 0.43 & 0.36  \\
 \hline
 Experiment~\cite{Stone:2014wza,Elliott:2013pna}   & --   & --   & $250-315$ &
 		 	$17.9 \pm 0.4$ & $0.06 \pm 0.01$ & $0.31 \pm 0.07$ & $0.29$ \\
\hline
\hline
 \end{tabular*}
\label{tab:comp}
\end{table*}

The situation regarding the eigenvolume parameter $b$ is somewhat more complicated. The modifications to the VDW attraction term  lead to smaller values of $b$. On the other hand, the CS modification of the repulsive term leads to larger values of $b$. The resulting range of values obtained in this work is $b=2.50-4.43$~fm$^3$. 

Whenever the EV corrections are used in the hadronic physics applications, 
the corresponding hard-core radius of a hadron is discussed.
Normally, the classical relation $b = 16 \pi r^3 /3$ between the hard-core radius and the eigenvolume parameter is used for this purpose. For nucleons, the values in the range $r=0.3-0.8$~fm had been reported in the literature~\cite{Yen:1997rv,Yen:1998pa,BraunMunzinger:1999qy,Cleymans:2005xv,Andronic:2005yp,Andronic:2012ut,Begun:2012rf}. The range $b=2.50 - 4.43$~fm$^3$ obtained in the present work corresponds to $r = 0.53 - 0.64$~fm.
The values in this range are notably higher than the $r \simeq 0.3$~fm value, estimated for a nucleon from the $NN$-scattering data, 
and used in, e.g., in Refs.~\cite{BraunMunzinger:1999qy,Andronic:2005yp,Andronic:2012ut}.
One has to keep in mind, however, that the above relation between $b$ and $r$ is inherently \emph{classical} one, without any quantum-mechanical effects.
The quantum mechanical calculation of the 2nd virial coefficient for systems with the hard-core repulsion~\cite{Kostyuk:2000nx,Typel:2016srf} suggests larger values of the parameter $b$ at lower temperatures. The classical limit, on the other hand, is only reached at very high temperatures. For nucleons, these temperatures may be much higher than those in the nuclear matter region. In that respect, the $r$ value used in the classical relation should be regarded only as an \emph{effective} hard-core radius. 
It should also be noted that the effects of the many-body interactions are expected to be important at the normal nuclear density. It is feasible that ``large'' values of $r$, extracted from the properties of the nuclear ground state, mimic some of these effects.
The fits to the nuclear ground state performed in this work appear to disfavor the values of the effective hard-core radius of nucleon smaller than 0.5~fm, at least for the nuclear matter region of the phase diagram. 
Still, it is possible that further modifications to the models
may lead to smaller values of $r$.

\subsection{Phase diagram and the liquid-gas transition}

\begin{figure}[t]
\centering
\includegraphics[width=0.49\textwidth]{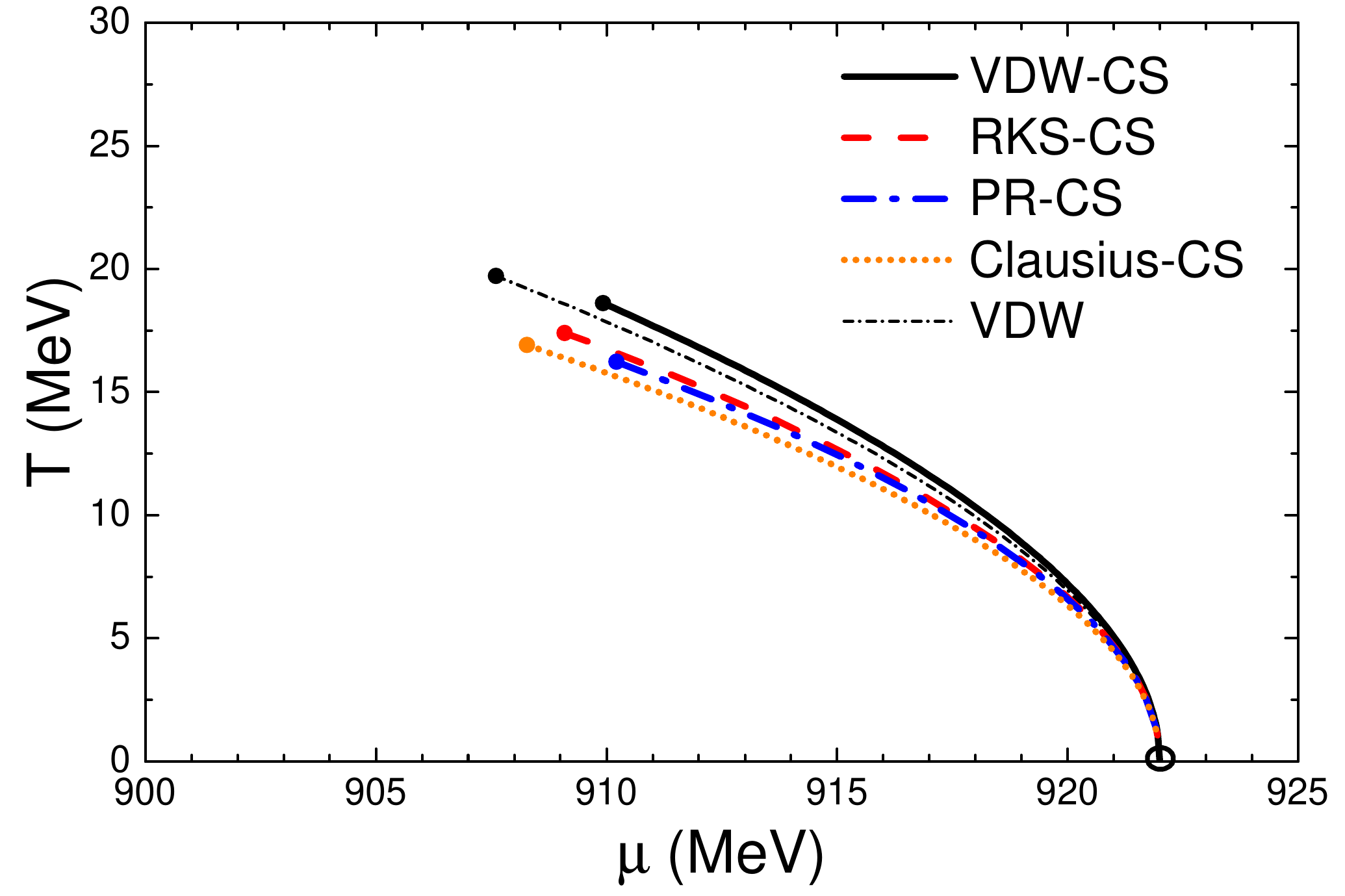}
\caption[]{
The lines of the first-order liquid-gas phase transition in the $T$-$\mu$ plane calculated within the VDW (black dash-dotted thin line), VDW-CS (solid black line), RKS-CS (red dashed line), PR-CS (dash-dotted blue line), and Clausius-CS (dotted orange line) models. The open circle corresponds to the nuclear ground state while the locations of the critical point in different models are denoted by the  solid colored circles.
}\label{fig:TMu-PT}
\end{figure}

\begin{figure}[t]
\centering
\includegraphics[width=0.49\textwidth]{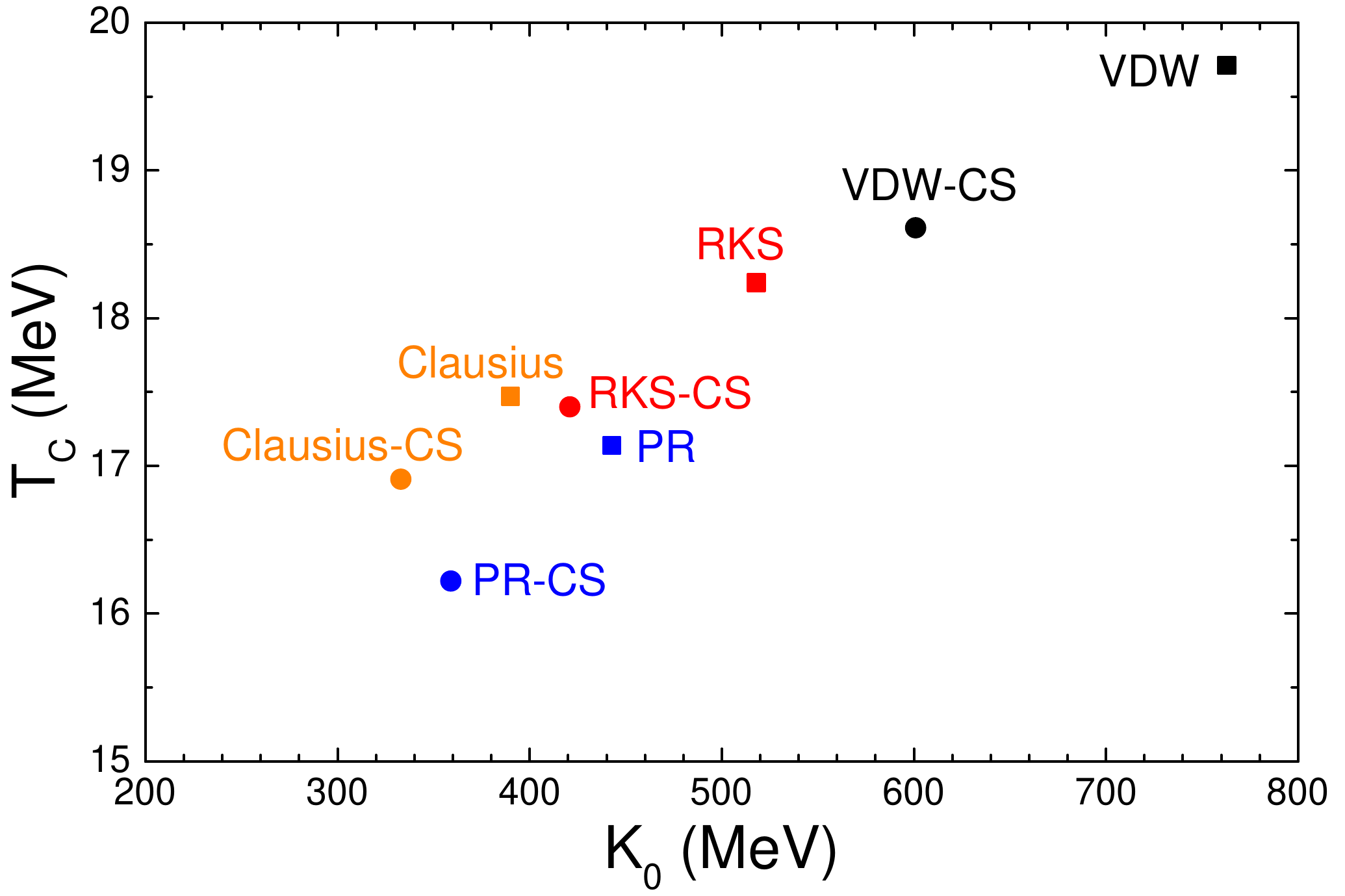}
\caption[]{
The correlation between the values of the critical temperature $T_c$ of the nuclear liquid-gas transition and the incompressibility factor $K_0$ within the real gas models.
}\label{fig:Tc-K0}
\end{figure}

All of the considered models predict the presence of the first-order nuclear liquid-gas phase transition which ends at the CP. The model predictions for the critical temperature $T_c$, the critical density $n_c$, the critical pressure $p_c$, and the critical compressibility ratio $Z_c = p_c / (n_c \, T_c)$, are listed in Table~\ref{tab:comp}. Also listed are the predictions of the Walecka model (taken from \cite{Silva:2008zza}), as well as the recent experimental estimates of Ref.~\cite{Elliott:2013pna}.
The phase transition lines $\mu = \mu_{\rm mix} (T)$
are exhibited in Fig.~\ref{fig:TMu-PT}. For the clarity of presentation, some of the models are omitted from Fig.~\ref{fig:TMu-PT}. 
For each model, the line starts from the nuclear ground state with $T = 0$ and $\mu_0 = 922$~MeV, and it ends at the CP. At each point of the phase transition line, two solutions with different particle densities and with equal pressures exist. These two solutions correspond, respectively, to the gas and to the liquid phase.
All of the considered models give a rather similar picture for the liquid-gas phase transition. The critical temperature varies in the range $T_c = 16.2 - 19.7$~MeV, with the highest value achieved in the VDW model. The critical density values are in the range 0.06-0.07~fm$^{-3}$. The values of $T_c$ and $n_c$ are in a reasonably good agreement with the recent experimental estimates of Ref.~\cite{Elliott:2013pna}. The critical pressure appears to be rather notably overestimated by the VDW model, which yields $p_c = 0.52$~MeV/fm$^3$. This value is in contrast to the experimental estimate of $p_c = 0.31 \pm 0.07$~MeV/fm$^3$. The Clausius-CS model, on the other hand, yields $p_c = 0.30$~MeV/fm$^3$, which is in a good agreement with this estimate. It is notable that the Clausius-CS model also yields the best description of the nuclear incompressibility factor among the considered models.

The values in Table~\ref{tab:comp} clearly suggest a presence of the correlation between the critical parameters $T_c$, $n_c$, $p_c$ and the incompressibility factor $K_0$. Higher values of $T_c$, $n_c$, and $p_c$ generally correspond to higher values of the $K_0$. This trend is illustrated in Fig.~\ref{fig:Tc-K0}, where the correlation between $T_c$ and $K_0$ is plotted. Our result is in line with the findings reported in Refs.\cite{Kapusta:1984ij,Lattimer:1991nc,Natowitz:2002nw,Rios:2010jh}, and, more recently, in Ref.~\cite{Lourenco:2016jdo}, where a much more systematic study on this subject within the relativistic mean field models was presented.

\subsection{Nucleon number fluctuations}
In the GCE, the number of particles fluctuates between different microscopic states, and only the average number can be fixed. These fluctuations, especially their higher order moments, provide an information about the finer details of the equation of state. Typically, they are characterized by the following dimensionless cumulants (susceptibilities),
\eq{\label{eq:chin}
\chi_n = \frac{\partial^n (p/T^4)}{\partial(\mu/T)^n}~.
}

\begin{figure*}[t]
\centering
\includegraphics[width=0.49\textwidth]{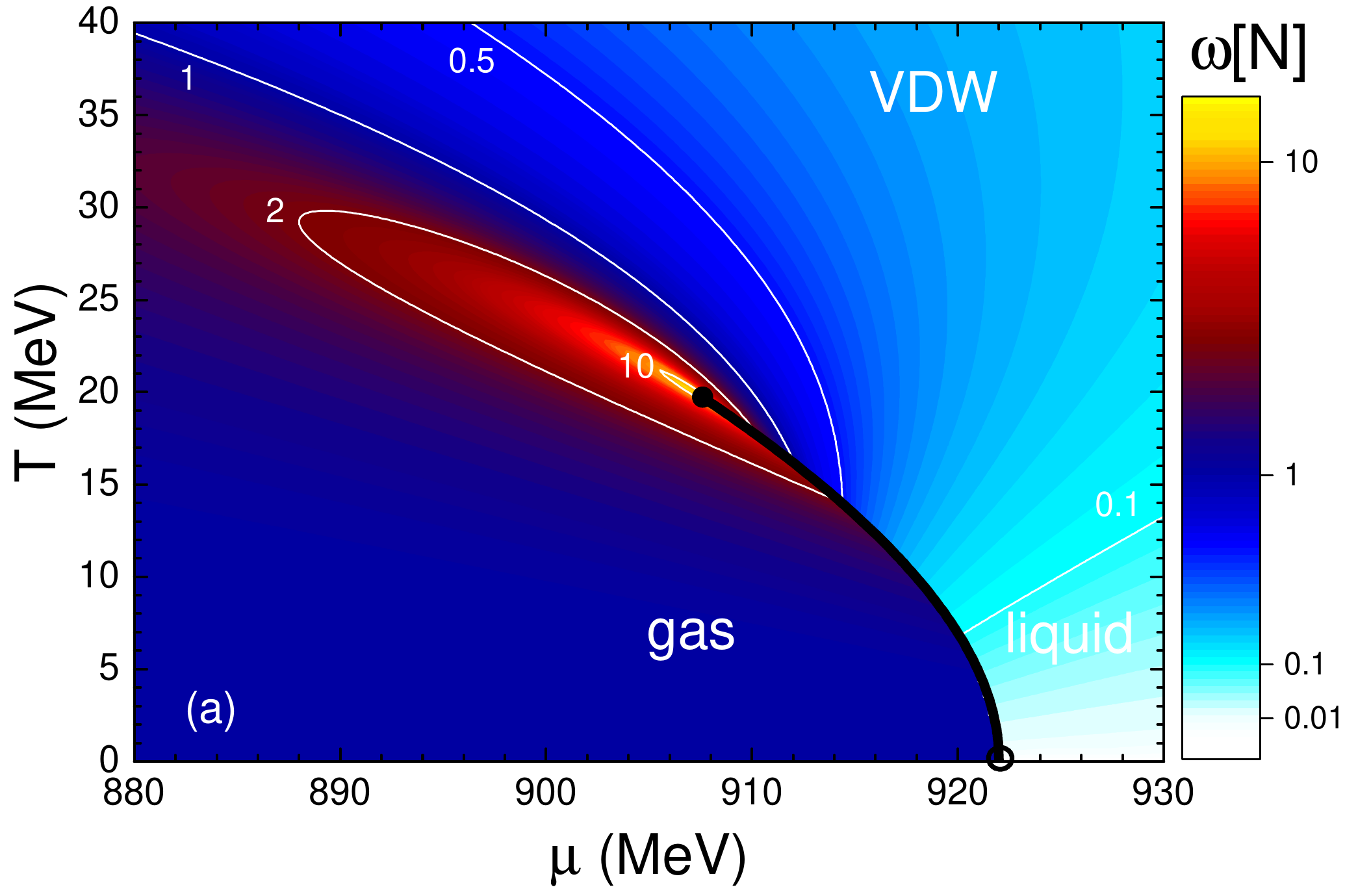}
\includegraphics[width=0.49\textwidth]{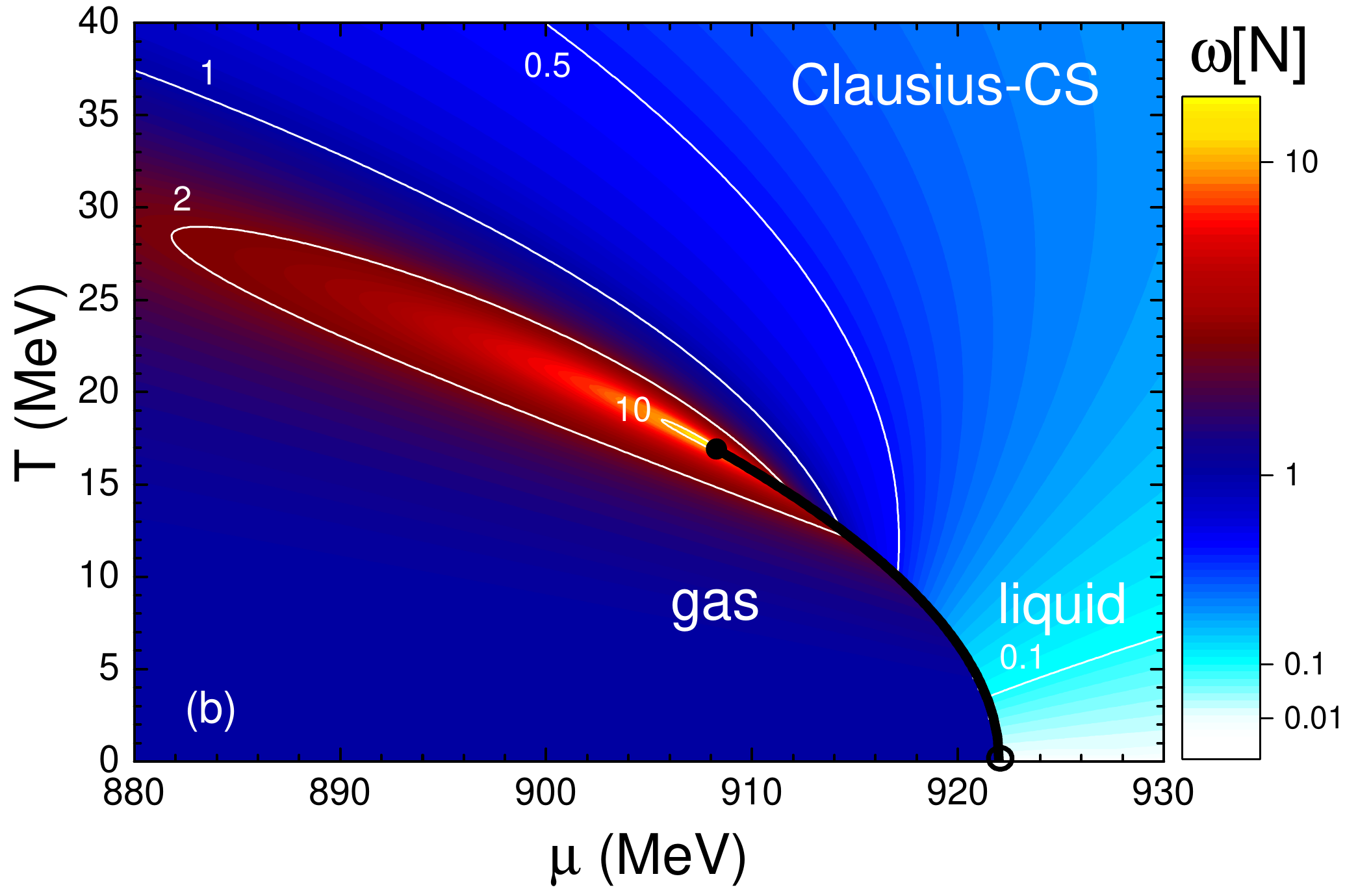}
\includegraphics[width=0.49\textwidth]{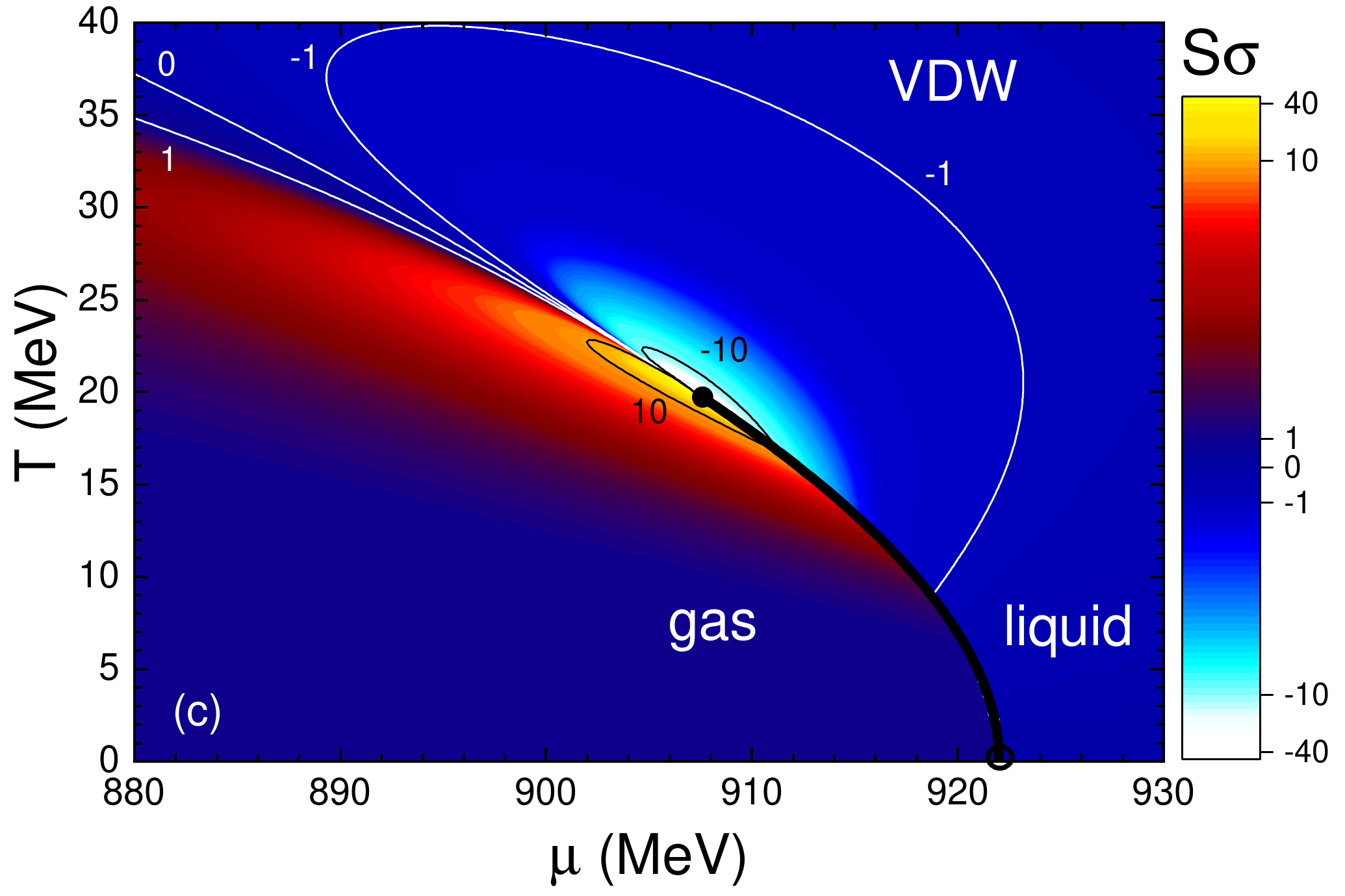}
\includegraphics[width=0.49\textwidth]{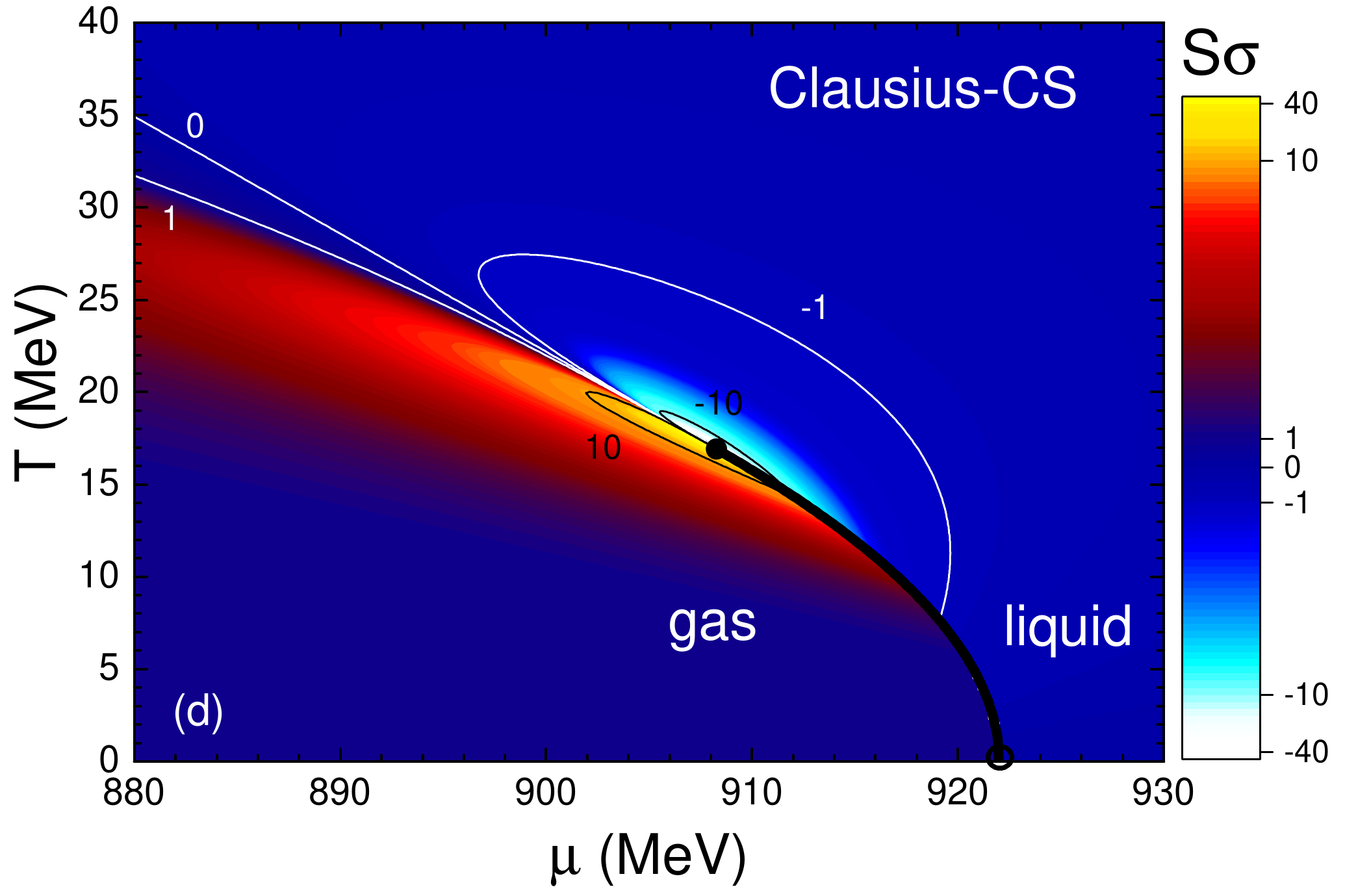}
\includegraphics[width=0.49\textwidth]{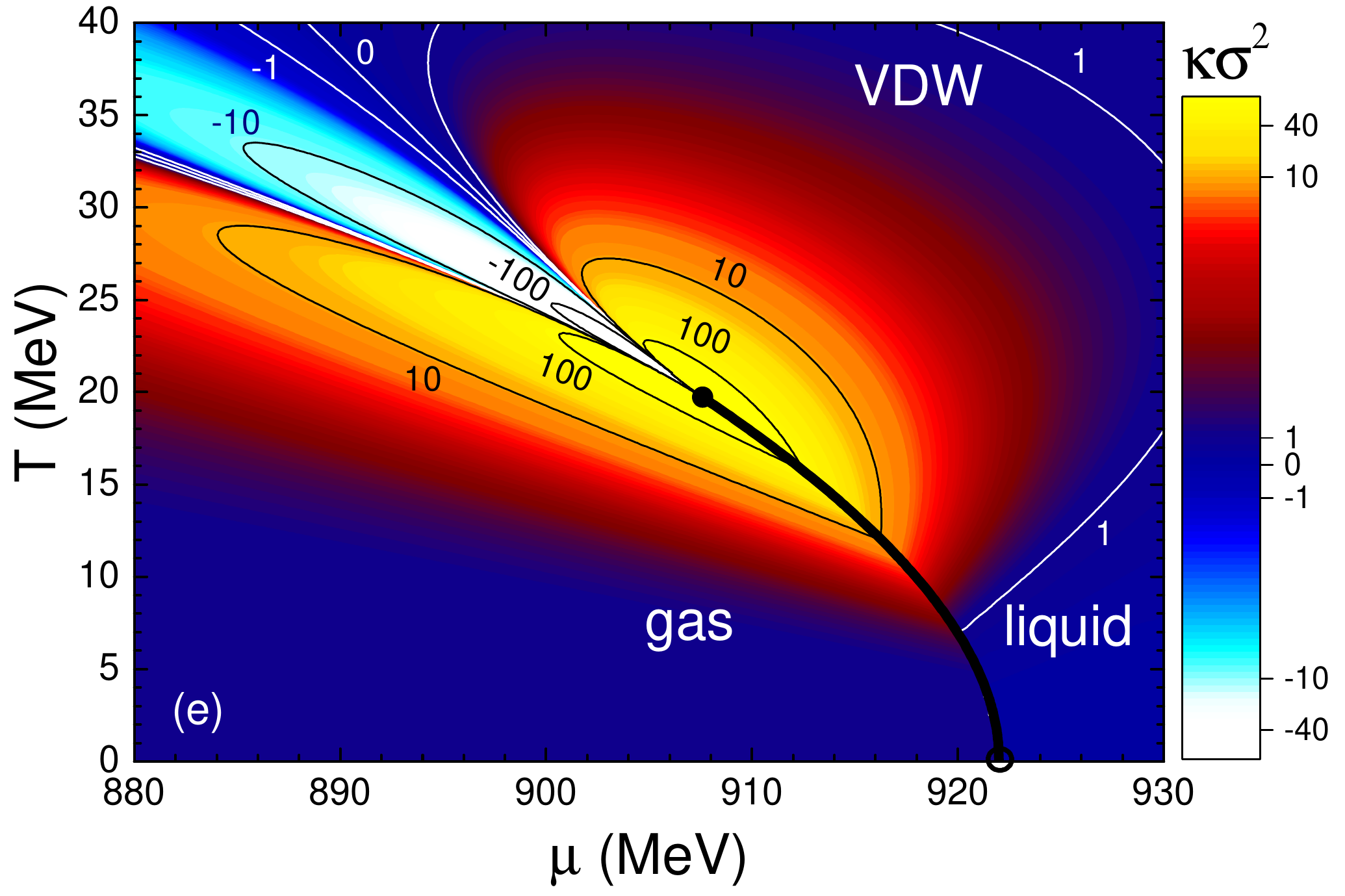}
\includegraphics[width=0.49\textwidth]{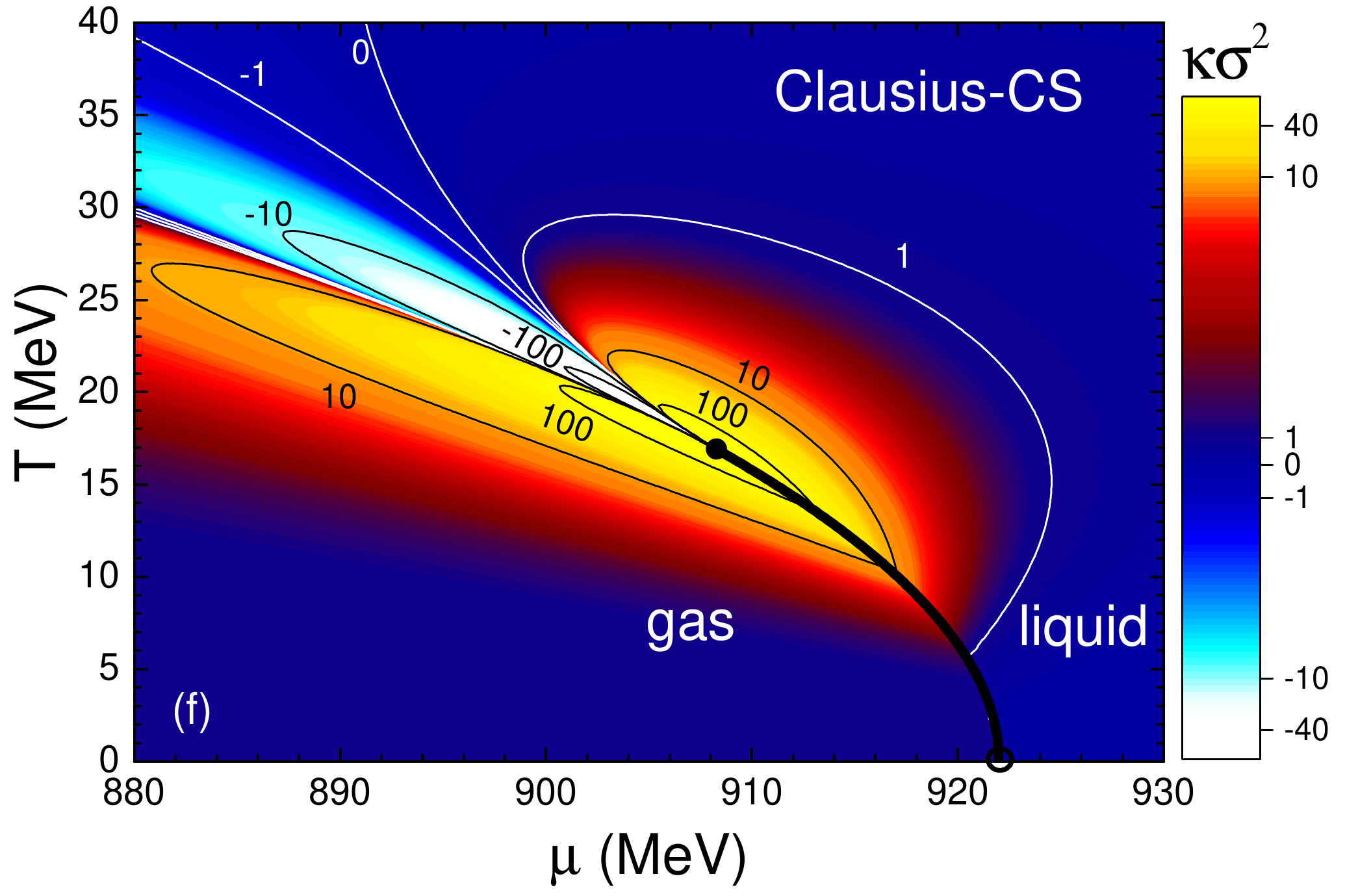}
\caption[]{
%
The contour plots for (a)-(b) scaled variance
$\omega[N] = \chi_2 / \chi_1$, 
(c)-(d) skewness
$S \sigma = \chi_3 / \chi_2$,
and (e)-(f) kurtosis
$\kappa \sigma^2 = \chi_4 / \chi_2$
calculated for the symmetric nuclear matter
in $(T,\mu)$ coordinates within the VDW equation of state (left panels) and the Clausius-CS equation of state (right panels).
The light blue (white) colors correspond to the regions where the scaled variance is small, $\omega[N] \ll 1$, and where the $S \sigma$ and $\kappa \sigma^2$ attain significant negative values.
The dark blue (black) colors correspond to the regions where the scaled variance, skewness, and kurtosis are close to the Poisson expectation, i.e. $\omega[N] \simeq 1$, $S \sigma \simeq 1$, and $\kappa \sigma^2 \simeq 1$.
The red and yellow (gray) colors correspond to the regions where the scaled variance is large $\omega[N] \gg 1$, and where the $S \sigma$ and $\kappa \sigma^2$ attain significant positive values.
The open circle at $T=0$ denotes the ground state of nuclear matter,
the solid circle at $T=T_c$ corresponds to the CP,
and the phase transition curve $\mu=\mu_{\rm mix}(T)$ is depicted by the thick solid line.
}\label{fig:Tmu-fluc}
\end{figure*}


Studies of the event-by-event fluctuations in high energy nucleus-nucleus collisions are presently actively being used to uncover the properties of the strongly interacting matter~(see, e.g. Refs.~\cite{Koch:2008ia,Stephanov:1999zu,Gorenstein:2015ria}), in particular, in the context of the search for the QCD critical point~\cite{Stephanov:1998dy,Gazdzicki:2015ska}. The higher-order (non-Gaussian)  measures of the fluctuations of conserved charges were suggested to study the QCD phase structure. Experimentally, the
search for the CP is in progress. The higher moments of the net-proton and net-charge multiplicity distributions in Au+Au collisions were recently measured by the STAR collaboration for the $\sqrt{s_{_{\rm NN}}} = 7.7-200$~GeV energy range~
\cite{Aggarwal:2010wy,Adamczyk:2013dal,Adamczyk:2014fia}. However, no definitive conclusion regarding the existence and location of the QCD CP has been obtained yet.


The following volume-independent cumulant ratios are particularly useful
\eq{
\omega[N] & = \frac{\chi_2}{\chi_1} =  \frac{\langle (\Delta N)^2\rangle}{\langle N\rangle}~, 
\\
S\sigma & =  \frac{\chi_3}{\chi_2} =  \frac{\langle (\Delta N)^3\rangle}{\langle (\Delta N)^2\rangle}~,
\label{N}
\\
\kappa\sigma^2 & =  \frac{\chi_4}{\chi_2} =  \frac{\langle (\Delta N)^4\rangle - 3\langle (\Delta N)^2\rangle^2}{\langle (\Delta N)^2\rangle}~,
}
where $\Delta N\equiv N-\langle N\rangle$, the
$\langle (\Delta N)^2\rangle$,  $\langle (\Delta N)^3\rangle$, and
$\langle (\Delta N)^4\rangle$ are the central moments, and where $\langle \ldots\rangle$
denotes the ensemble averaging.
These cumulant ratios are called the scaled variance $\omega[N]$, skewness $S \sigma$, and kurtosis $\kappa \sigma^2$. Their behavior in the vicinity of the CP was calculated in various effective QCD models~\cite{Stephanov:2011pb,Schaefer:2011ex,Chen:2015dra}. It is understood that the behavior of these quantities in the vicinity of the CP is governed by the universality. This implies that qualitatively this behavior should be the same for all models with a CP which belong to the same universality class. The critical behavior in nuclear matter, in particular concerning the critical exponents, was previously studied in Refs.~\cite{Silva:2008zza,Dutra:2008zz,Rios:2010jh}.


Lately, these fluctuation measures were calculated for the symmetric nuclear matter within the VDW equation with Fermi statistics for nucleons~\cite{Vovchenko:2015pya}. Qualitative features of the critical fluctuations, obtained in that work, are consistent with effective QCD models as well as with the
analytic results of the classical VDW equation~\cite{Vovchenko:2015uda}.
It is, however, instructive to investigate the robustness of these results by considering different models of the symmetric nuclear matter. Present results already show very different possibilities for the values of the incompressibility factor $K_0$ between different real gas models. Will a similarly sensitive behavior be seen in the cumulants?

The GCE formulation of the real gas models~(Sec.~\ref{sec-full})
allows to calculate the 2nd and higher order cumulants of the particle number fluctuations.
This section shows the results for $\omega[N]$, $S \sigma$, and $\kappa \sigma^2$, obtained within two representative models: the VDW model and the Clausius-CS model. Results obtained within the other considered models lie roughly ``in-between'' these two.

The results of the calculations for the scaled variance $\omega[N] = \chi_2 / \chi_1$, skewness $S \sigma$, and kurtosis $\kappa \sigma^2$ are exhibited in Fig.~\ref{fig:Tmu-fluc}. 
The two models give qualitatively similar fluctuation patterns for all three observables. 
Since these were previously considered in great details in Ref.~\cite{Vovchenko:2015pya}, their description is not repeated here. These fluctuation patterns near the CP are consistent with the model-independent universality arguments regarding the critical behavior in the vicinity of the QCD critical point~\cite{Stephanov:2008qz,Stephanov:2011pb}. The patterns are also fully consistent with the analytic predictions of the classical VDW equation~\cite{Vovchenko:2015uda}. The similarity of the
results between the VDW and Clausius-CS models illustrates the universality of the critical behavior with actual
quantitative model calculations.

The biggest quantitative differences are observed in the liquid phase at low temperatures, and also in the crossover region above the CP. 
These differences are more pronounced for higher moments.
Thus, the variations in EV and mean-field mechanisms are important mainly for the liquid phase. 

\section{From nuclear matter to HRG}
\label{sec-RGHRG}

The real gas models can describe some of the basic properties of the nuclear matter. 
The simplicity of the approach, however, does no justice to the complexity of the many-body nucleon interactions.
The mean field models, based on the relativistic mean-field theory or on the Skyrme approach, are likely preferable for a better quantitative description of the many involved nuclear matter properties~(see~\cite{Dutra:2012mb,Dutra:2014qga} for an overview).
On the other hand, the present approach allows a straightforward generalization to a multi-component hadron gas. This, in turn, opens new applications in the physics of heavy-ion collisions and QCD equation of state.

\begin{figure*}[t]
\begin{center}
\includegraphics[width=0.49\textwidth]{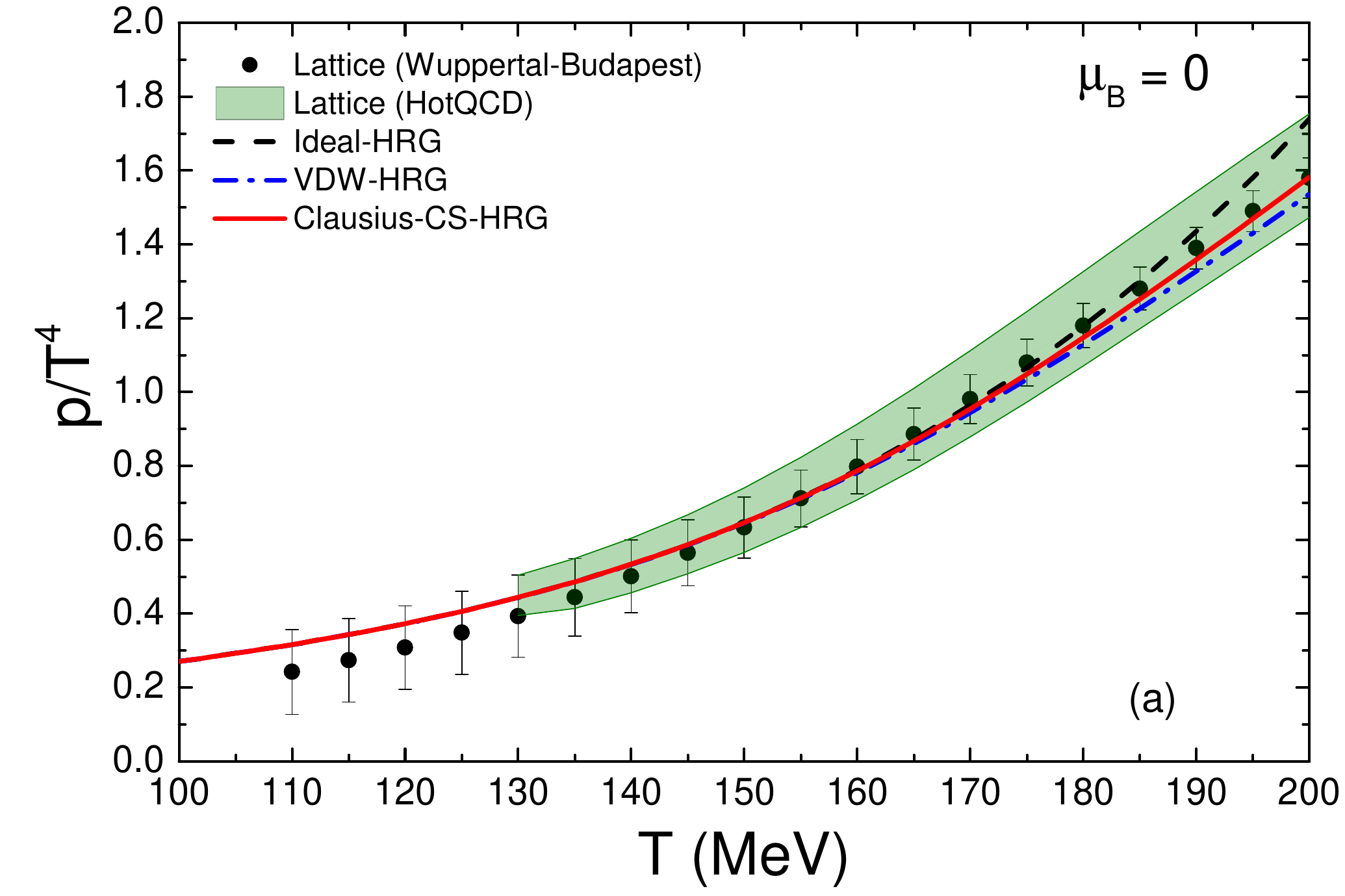}
\includegraphics[width=0.49\textwidth]{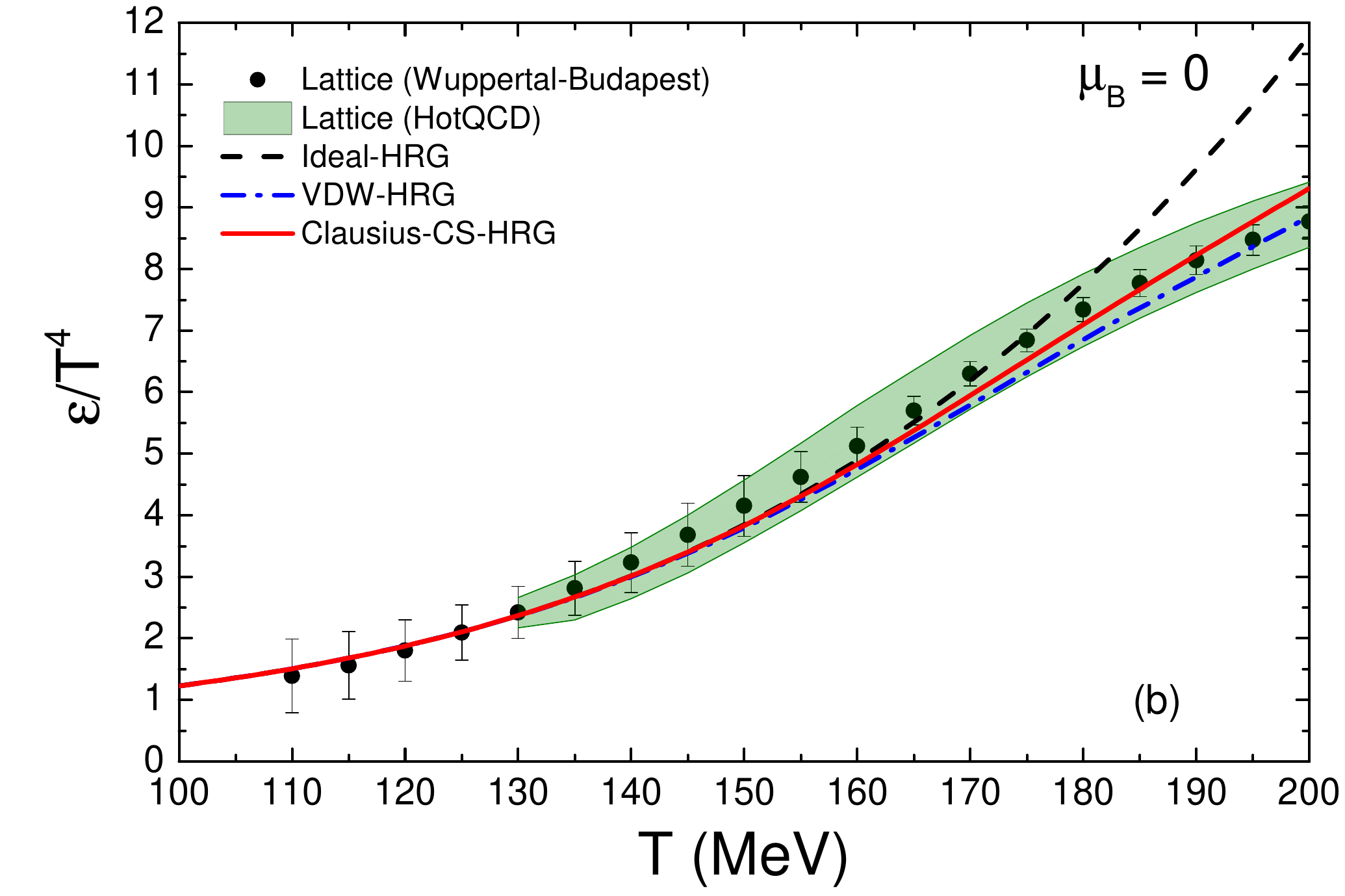}
\includegraphics[width=0.49\textwidth]{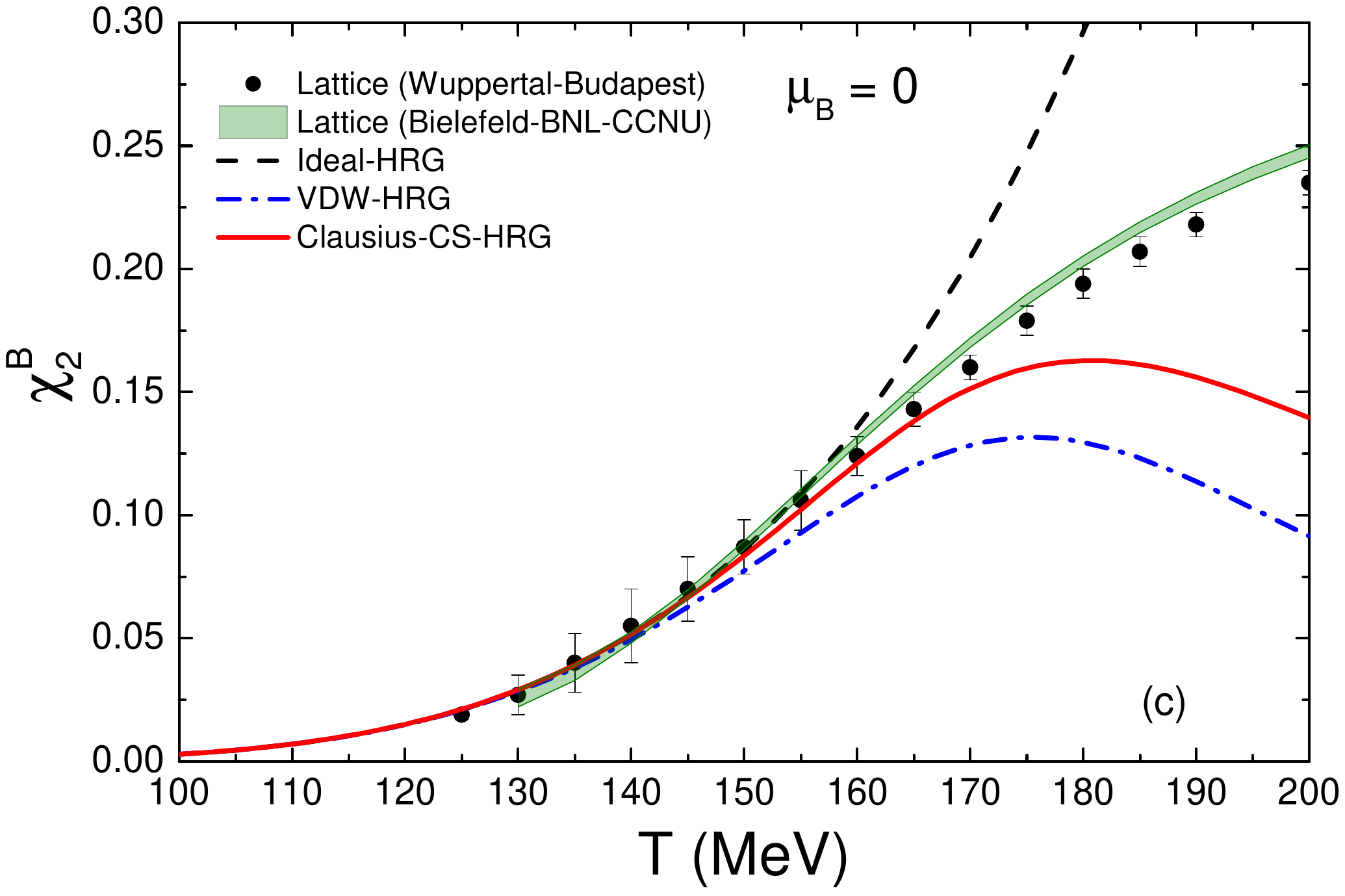}
\includegraphics[width=0.49\textwidth]{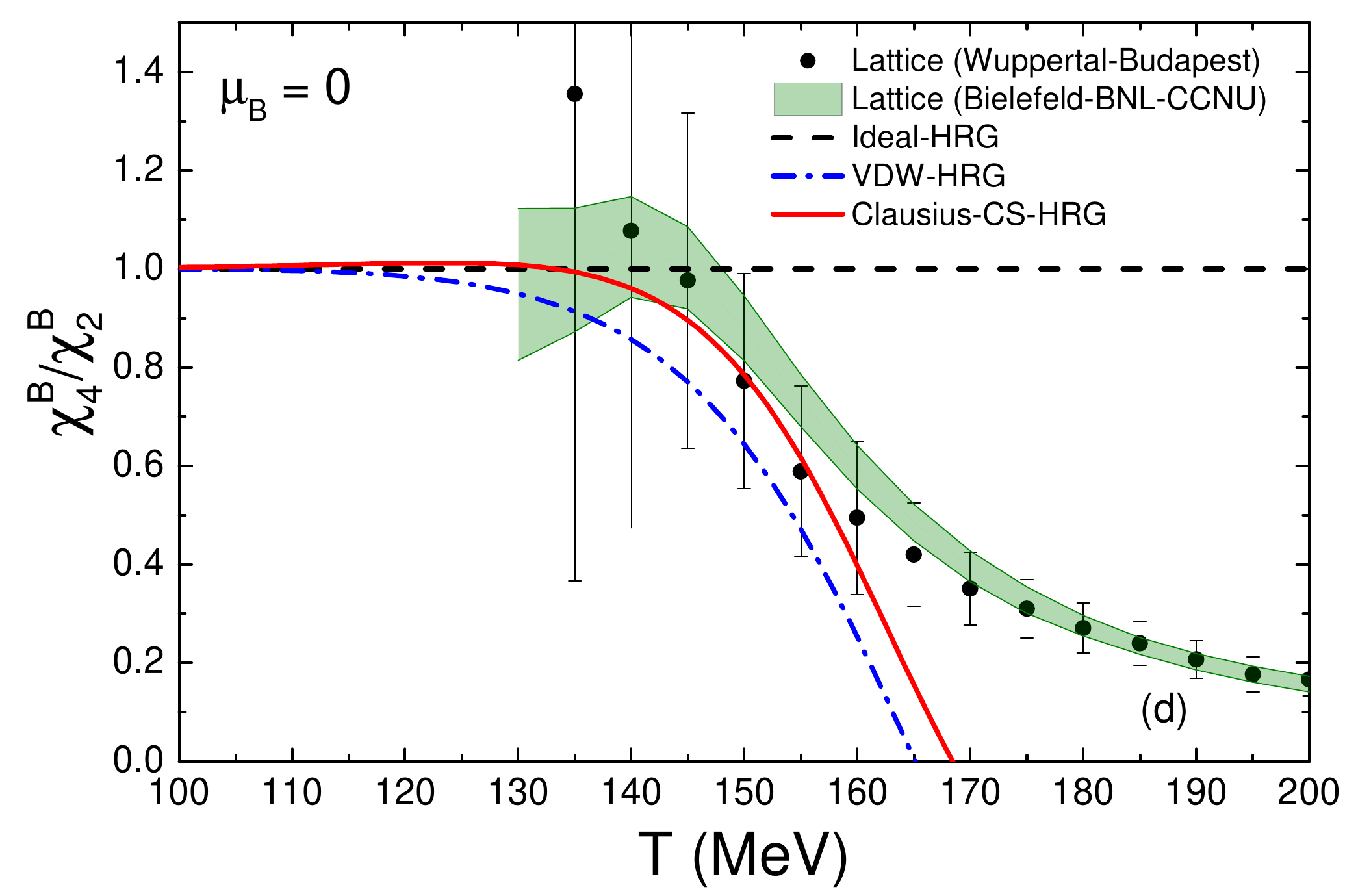}
\caption{
The temperature dependence of (a) the scaled pressure $p/T^4$, (b) the scaled energy density $\varepsilon/T^4$, and the net baryon number susceptibilities (c) $\chi_2^B$ and (d) $\chi_4^B/\chi_2^B$.
Calculations are done within the Ideal-HRG model (dashed black lines), the VDW-HRG model (dash-dotted blue lines), and the Clausius-CS-HRG model (solid red lines). 
The parameters $a$ and $b$ for the VDW-HRG and Clausius-CS-HRG models are listed in Table~\ref{tab:comp}.
The recent lattice QCD results of the Wuppertal-Budapest~\cite{Borsanyi:2013bia,Bellwied:2013cta,Bellwied:2015lba} 
and the HotQCD/Bielefeld-BNL-CCNU~\cite{Bazavov:2014pvz,Bazavov:2017dus} collaborations are shown, respectively, by symbols and green bands.
}
\label{fig:RGvsLattice}
\end{center}
\end{figure*}

To illustrate the latter point, let us consider a simple generalization of the ideal HRG model which allows to include the VDW-like interactions between baryons in the framework of a real gas model. Following Ref.~\cite{Vovchenko:2016rkn}, it is assumed that the baryon-baryon and antibaryon-antibaryon interactions are the same as the nucleon-nucleon interaction.
At the same time, the baryon-antibaryon, meson-baryon, and meson-meson interactions are explicitly omitted.
The resulting model consists of three independent
sub-systems: Non-interacting mesons, interacting baryons, and interacting antibaryons. Due to the assumed similarity between baryon-baryon and nucleon-nucleon interaction, the real gas models can be straightforwardly generalized to the case of a multi-component HRG.
The total pressure reads
\begin{equation}
p(T,\bs \mu) = p_M(T,\bs \mu) + p_B(T,\bs \mu) + p_{\bar{B}}(T,\bs \mu),
\end{equation}
with
\begin{align}
p_M(T,\bs \mu) & =
\sum_{j \in M} p_{j}^{\rm id} (T, \mu_j) \\
\label{eq:PB}
p_B(T,\bs \mu) & =
[f(\eta_B) - \eta_B\, f'(\eta_B)] \, \sum_{j \in B} p_{j}^{\rm id} (T, \mu_j^{B*}) \nonumber \\
& \quad + n_B^2 \, u'(n_B), \\
\label{eq:PBBar}
p_{\bar{B}}(T,\bs \mu) & =
[f(\eta_{\bar{B}}) - \eta_{\bar{B}}\, f'(\eta_{\bar{B}})] \, \sum_{j \in \bar{B}} p_{j}^{\rm id} (T, \mu_j^{\bar{B}*}) \nonumber \\
& \quad + n_{\bar{B}}^2 \, u'(n_{\bar{B}}),
\end{align}
where $M$ stands for mesons, $B$ for baryons, and $\bar{B}$ for antibaryons, $p_{j}^{\rm id}$ is the Fermi or Bose ideal gas pressure,
$\bs \mu=(\mu_B,\mu_S,\mu_Q)$ are the chemical potentials which regulate the average values of the net baryon number $B$, strangeness $S$, electric charge $Q$,
and 
\eq{
n_{B(\bar{B})} = \sum_{i \in B(\bar{B})} n_i, 
\quad \textrm{and} \quad
\eta_{B(\bar{B})} = \frac{b}{4} \, \sum_{i \in B(\bar{B})} n_i,
}
are, respectively, the total density and the packing fraction of all (anti)baryons.
The total density of baryons, $n_B$, satisfies the equation
\eq{
n_B = f(\eta_B) \, \sum_{i \in B(\bar{B})} n_{i}^{\rm id} (T, \mu_i^{\bar{B}*})
}
and the shifted chemical potentials, $\mu_i^{B*}$, are given by
\eq{\label{eq:dmuB}
\mu_i^{B*} - \mu_i & = \Delta \mu_B 
\nonumber \\
& = \frac{b}{4} \, f'(\eta_B) \, \sum_{j \in B}  p_{j}^{\rm id} (T, \mu_j^{B*})
\nonumber \\
& \quad - u(n_B) - n_B u'(n_B).
}

The numerical solution to Eq.~\eqref{eq:dmuB} allows to obtain $\Delta \mu_B$. This allows to calculate all other
quantities in the baryon subsystem. The same procedure is applied for the antibaryon subsystem. Calculations in the mesonic sector are straightforward. The list of hadrons included in the HRG model includes all strange and non-strange hadrons which are listed in the Particle Data Tables~\cite{Agashe:2014kda} and have a confirmed status there~(see~\cite{Vovchenko:2016rkn} for more details).

Clearly, the model is rather simplistic. Nevertheless, it does allow to include the essential features of the nuclear matter physics into the hadronic equation of state. Unlike Ref.~\cite{Vovchenko:2016rkn}, where the nuclear matter description was restricted to the standard VDW model, here an arbitrary EV fraction, $f(\eta)$, and an arbitrary density-dependent mean-field, $u(n)$, are allowed.
This allows to study possible correlations between the behavior of the HRG observables and the nuclear matter properties.
The thermodynamic properties of the HRG at zero chemical potential are considered as an example.

In this section, calculations are performed within two models: the VDW-HRG model, where baryon-baryon interactions are described by the standard VDW equation, and the Clausius-CS-HRG model, where the EV interactions are described by the CS model and where the attraction is taken in the Clausius form. 
The parameters $a$ and $b$ for both models were already fixed by the nuclear ground state properties in the previous section, and their values are listed in Table~\ref{tab:comp}. 
The calculations performed within the standard ideal HRG model are also shown for completeness.

\begin{figure}[t]
\centering
\includegraphics[width=0.49\textwidth]{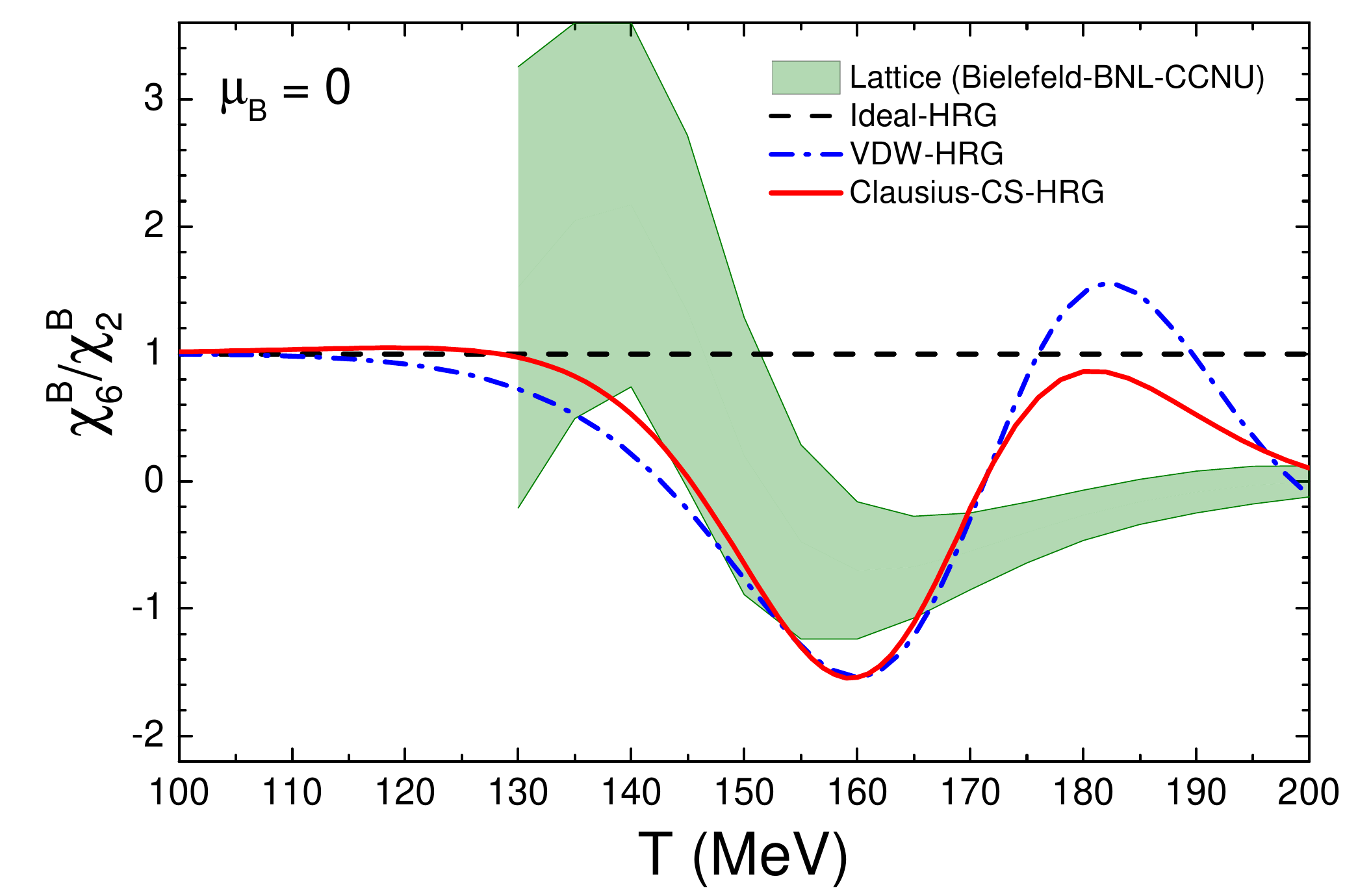}
\caption[]{
The temperature dependence of the $\chi_6^B/\chi_2^B$ susceptibility ratio for the net baryon number fluctuations.
Calculations are done within the Ideal-HRG model (dashed black lines), the VDW-HRG model (dash-dotted blue lines), and the Clausius-CS-HRG model (solid red lines). 
The parameters $a$ and $b$ for the VDW-HRG and Clausius-CS-HRG models are
listed in Table~\ref{tab:comp}.
The recent lattice QCD results of the Bielefeld-BNL-CCNU collaboration~\cite{Bazavov:2017dus} are shown by the green band.
}\label{fig:chi6}
\end{figure}

The temperature dependencies of the scaled pressure $p/T^4$ and the scaled energy density $\varepsilon/T^4$ are shown in the upper panels of Fig.~\ref{fig:RGvsLattice}. The calculations are performed at zero chemical potentials, i.e. $\mu_B = \mu_Q = \mu_S = 0$.
These results are compared to the lattice QCD data of the Wuppertal-Budapest~\cite{Borsanyi:2013bia} and of the HotQCD~\cite{Bazavov:2014pvz} collaborations. 
The inclusion of the baryon-baryon interaction terms leads to a modest suppression of the pressure and of the energy density at high temperatures $T \gtrsim 175$~MeV. The result is quite similar in both, VDW-HRG and Clausius-CS-HRG models.
This suppression slightly improves the agreement with the lattice data.
However, the pressure and the energy density are not very sensitive to the details of baryon-baryon interactions, especially at the lower temperatures. This is not surprising: the matter is meson-dominated at $\mu_B = 0$, and the mesonic contributions dominate over the baryonic ones for these two observables.

On the other hand, the baryon-related observables, such as the net-baryon number fluctuations (susceptibilities), can certainly be expected to be more sensitive to the modeling of the baryonic interactions. The $i$th order susceptibility is defined as
$\chi_i^B = \partial^i (p/T^4) / \partial(\mu_B/T)^i$. Due to the baryon-antibaryon symmetry, only the even order susceptibilities are non-zero at $\mu_B = 0$. 
The calculations for the $\chi_2^B$ and for the kurtosis ratio $\chi_4^B / \chi_2^B$ are shown in the lower panels of Fig.~\ref{fig:RGvsLattice}.
The baryon-baryon interactions have strong effect on both observables.
Both the VDW-HRG and Clausius-CS-HRG models predict an inflection point in the temperature dependence of $\chi_2^B$. This is at odds with the ideal HRG model, but consistent with the lattice data. 
The quantitative agreement with the lattice data in the crossover region ($T \sim 140-170$~MeV) is much better in the Clausius-CS-HRG model than in the VDW-HRG model or in the ideal HRG model. 
This is an interesting observation since the Clausius-CS-HRG model also leads to a much improved description of the nuclear matter incompressibility: $K_0$ of 333 MeV in the Clausius-CS-HRG model versus 762~MeV in the VDW-HRG model. Similarly, a drop of the $\chi_4^B / \chi_2^B$ ratio, seen in lattice simulations, is better reproduced by the Clausius-CS-HRG model than by other considered models.

The lattice QCD calculations for the sixth-order $\chi_6^B / \chi_2^B$ cumulant ratio for the
net baryon number fluctuations had recently been presented by the Bielefeld-BNL-CCNU collaboration~\cite{Bazavov:2017dus}. 
This ratio shows a non-monotonic temperature dependence in the crossover region~(Fig.~\ref{fig:chi6}). A non-monotonic dependence is also predicted by both VDW-HRG and Clausius-CS-HRG models. The lattice data is described better by the  Clausius-CS-HRG model.

These results show that the behavior of the baryon-related lattice QCD observables correlates strongly with the properties of nuclear matter: a model with an improved description of $K_0$ also leads to an improved description of the temperature dependence of $\chi_i^B$ at zero chemical potential.

\section{Summary}
\label{sec-summary}
The present work introduces a formalism to augment the classical models of equation of state for real gases with the quantum statistical effects.
This formalism can be generalized in a relatively straightforward manner to the multi-component systems.

A distinct novel feature of this work is the quantum statistical generalization of an arbitrary EV model.
In particular, the present formalism generalizes the Carnahan-Starling EV model to include the quantum statistics. 
The importance of the missing quantum statistical effects in the Carnahan-Starling model was previously pointed out in Ref.~\cite{Satarov:2014voa}.
The Carnahan-Starling model allows to improve the description of the nuclear matter properties, and of the lattice QCD data.

The quantum statistical real gas models
are shown to give a fairly
good description of the basic properties of the symmetric nuclear matter.
A family of only two-parameter VDW-like models is considered in the present work,
where the short-range repulsive interactions are modeled with the EV correction while the attractive interactions are described by the density-dependent mean-field.
Variations on the repulsive EV term include the VDW and Carnahan-Starling models,
while the considered attractive 
terms
are based on the VDW, Redlich-Kwong-Soave, Peng-Robinson, and Clausius equations of state.
The combination of the two different EV terms with the four different attraction terms gives the total of eight VDW-like models under consideration.
All models are consistent with the VDW equation of state in the low-density limit. Thus, in the region of small densities, the parameters $a$ and $b$ in each model can really be regarded as the VDW parameters.

The VDW parameters of the nucleon-nucleon interaction are fitted in each model to the properties of the ground state of nuclear matter.
The following range of parameters is obtained: $a = 330 \div 430$~MeV~fm$^3$, $b = 2.5 \div 4.4$~fm$^3$.
Fits within the standard VDW model are found to underestimate the value of the attraction parameter $a$.
In the context of the EV approach,
the fits to the nuclear ground state disfavor values of the effective hard-core radius of nucleon smaller than $0.5$~fm,
at least in the nuclear matter region of the phase diagram.
This seems to be an important constraint as the EV approach is very often employed in the hadronic physics applications, in particular as an extension of the HRG model. 
Still, the possibility of the temperature dependent effective hard-core radius cannot be excluded.

Modifications to the standard VDW repulsion and attraction terms allow to improve significantly the value of the nuclear incompressibility factor $K_0$, bringing it closer to the empirical estimates. Among the considered equations of state, the Clausius model with the Carnahan-Starling repulsion term was found to give the lowest value of $K_0 = 333$~MeV, which yields the best agreement with the available empirical estimates.
Remarkably, this two-parameter Clausius-CS model gives a better description of the incompressibility than the two-parameter versions of the Walecka (linear $\sigma$-$\omega$) model~(see e.g. \cite{Dutra:2014qga} for a review).
The present formalism permits variation of the third parameter $c$ in the Clausius model.
The parameter $c$ can be varied in order to obtain the needed value of $K_0$.
Thus, the Clausius model leaves room for further improvement in the description of the nuclear matter properties. 
A correlation between the $K_0$ and the critical parameters $T_c$, $n_c$, and $p_c$ of the nuclear liquid-gas phase transition is observed: higher values of $K_0$ generally correspond to larger ($T_c$, $n_c$, $p_c$) values, in line with  previous results obtained within other models of the nuclear matter~\cite{Kapusta:1984ij,Lattimer:1991nc,Natowitz:2002nw,Rios:2010jh,Lourenco:2016jdo}.

The fluctuation signatures of the nuclear matter critical point, namely the scaled variance, skewness, and
kurtosis, are also calculated within the quantum statistical
real gas models.
All models show essentially the same qualitative behavior of the critical fluctuations.
The comparison between the models illustrates the universality of the critical behavior.

The main advantage of the proposed formalism is that the real gas approach permits a relatively straightforward generalization to the multi-component hadron gas. This opens new applications in the physics of the heavy-ion collisions, and of the QCD equation of state.
This is illustrated by the calculations within the HRG models with baryon-baryon interactions. The calculations show that the behavior of the baryon-related lattice QCD observables at zero chemical potential correlates strongly with the nuclear matter properties such as the nuclear incompressibility.
An improved description of the lattice data for the net baryon susceptibilities 
correlates with an improved description of the nuclear incompressibility factor $K_0$.
In particular, the Clausius-CS model is shown to be considerably better than the standard VDW model for the description of both, the nuclear matter properties, and the lattice QCD data.
Further studies in this direction are of interest.

\vspace{0.5cm}
\section*{Acknowledgements}
The author thanks D.~Anchishkin and M.~Gorenstein for careful reading of the manuscript and for useful remarks.
Many fruitful discussions with P.~Alba, C.~Greiner, L.~Satarov, and H.~Stoecker are gratefully acknowledged.
The author appreciates the support from HGS-HIRe for FAIR.

\begin{widetext}
\appendix

\section{Ideal gas relations}
\label{app-ig}

This Appendix lists various thermodynamic functions of the ideal quantum gas in the GCE. They are used in the calculations within the real gas models performed in this work. The pressure, the particle number density, and the energy density read
\eq{
p^{\rm id} (T, \mu) ~ & = ~\frac{d}{6 \pi^2} \int d k
\frac{k^4}{\sqrt{m^2 + k^2}} \, \left[ \exp\left(\frac{\sqrt{m^2+k^2}-\mu}{T}\right)
+ \alpha \right]^{-1}~, \\
n^{\rm id} (T, \mu) ~ & = ~\frac{d}{2 \pi^2} \int d k \,
k^2 \, \left[ \exp\left(\frac{\sqrt{m^2+k^2}-\mu}{T}\right)
+ \alpha \right]^{-1}~, \\
\varepsilon^{\rm id} (T, \mu) ~ & = ~\frac{d}{2 \pi^2} \int d k \,
k^2 \, \sqrt{m^2+k^2} \, \left[ \exp\left(\frac{\sqrt{m^2+k^2}-\mu}{T}\right)
+ \alpha \right]^{-1}~, \\
s^{\rm id} (T, \mu) ~ &= ~\frac{\varepsilon^{\rm id} (T, \mu) + p^{\rm id} (T, \mu) - \mu \, n^{\rm id} (T, \mu)}{T},
}
where $\alpha = +1$ for fermions (nucleons), $\alpha = -1$ for bosons, and $\alpha = 0$ for the Boltzmann approximation.

The $k$th moment $n_{(k)}^{\rm id} (T,\mu)$ of the 
occupation number is
\eq{
n_{(k)}^{\rm id} (T, \mu) ~ & = ~\frac{d}{2 \pi^2} \int d k \,
k^2 \, \left[ \exp\left(\frac{\sqrt{m^2+k^2}-\mu}{T}\right)
+ \alpha \right]^{-k}~.
}
where it is evident that $n_{(1)}^{\rm id} (T, \mu) \equiv n^{\rm id} (T,\mu)$.

The ideal gas cumulants (susceptibilities) $\chi_k^{\rm id} (T,\mu) = \partial^k (p^{\rm id}/T^4) / \partial (\mu/T)^k$ read
\eq{
\chi_1^{\rm id} (T,\mu) ~ & = \frac{n^{\rm id} (T,\mu)}{T^3},\\
\chi_2^{\rm id} (T,\mu) ~ & = \frac{n^{\rm id} (T,\mu) - \alpha \, n^{\rm id}_{(2)} (T,\mu)}{T^3},\\
\chi_3^{\rm id} (T,\mu) ~ & = \frac{n^{\rm id} (T,\mu) - 3 \alpha \, n^{\rm id}_{(2)} (T,\mu) + 2\,\alpha^2\,n^{\rm id}_{(3)} (T,\mu)}{T^3},\\
\chi_4^{\rm id} (T,\mu) ~ & = \frac{n^{\rm id} (T,\mu) - 7 \alpha \, n^{\rm id}_{(2)} (T,\mu) + 12\,\alpha^2\,n^{\rm id}_{(3)} (T,\mu) - 6\, \alpha^3 \, n^{\rm id}_{(4)} (T,\mu)}{T^3}.
}

\section{Particle number fluctuations in real gas models}
\label{app-fluct}
This Appendix describes the calculation procedure for the cumulants of the GCE particle number fluctuations in the real gas models. Fluctuation measures up to the 4th order are considered. The process can be continued in a systematic manner up to an arbitrarily high order.
The following derivation is valid for any thermodynamic system for which the free energy can be written in the form given by Eq.~\eqref{eq:Fquant}.

The cumulants $\chi_k (T,\mu) = \partial^k (p/T^4) / \partial (\mu/T)^k$ are related to the particle number density and its derivatives with respect to the chemical potential $\mu$
\eq{\label{eq:chideriv}
\chi_1 (T,\mu) = \frac{n}{T^3},  \quad
\chi_2 (T,\mu) = \frac{1}{T^2} \, \left(\frac{\partial n}{\partial \mu} \right), \quad \chi_3 (T,\mu) = \frac{1}{T} \, \left(\frac{\partial^2 n}{\partial \mu^2}\right), \quad \chi_4 (T,\mu) = \left(\frac{\partial^3 n}{\partial \mu^3}\right),
}
where all partial derivatives throughout this Appendix are taken at a constant temperature. Recall that
\eq{
n = f(\eta) \, n^*,
}
and
\eq{
\mu^*  = \mu + \frac{b}{4} \, f'(\eta) \, p^* - u(n) - n\, u'(n).
}
Taking the derivative of the above two equations with respect to $\mu$ one obtains the following system of equations for $(\partial n / \partial \mu)$ and $(\partial \mu^* / \partial \mu)$
\begin{alignat}{3}
\left[ 1 - \frac{b}{4} \, f'(\eta) \, n^* \right] &\left(\frac{\partial n}{\partial \mu}\right) \quad & - & & f(\eta) \, \chi_2^* \, T^2 \, & \left(\frac{\partial \mu^*}{\partial \mu}\right) = 0,  
\label{eq:c21}
\\
\left[ -\left(\frac{b}{4}\right)^2 \, f''(\eta) \, p^* + v'(n) \right] &\left(\frac{\partial n}{\partial \mu}\right) \quad & + & & \quad \left[ 1 - \frac{b}{4} \, f'(\eta) \, n^* \right] &\left(\frac{\partial \mu^*}{\partial \mu}\right) = 1,
\label{eq:c22}
\end{alignat}
where $v(n) \equiv u(n) + n \, u'(n)$ and $\chi_k^* \equiv \chi_k^{\rm id} (T, \mu^*)$. 

The system of equations \eqref{eq:c21} and \eqref{eq:c22} can be solved to obtain $(\partial n / \partial \mu)$ and $(\partial \mu^* / \partial \mu)$, which in turn allows one to calculate the $\chi_2$ from~\eqref{eq:chideriv}. The following closed-form expression for $\omega[N] = \chi_2 / \chi_1$ is obtained
\eq{\label{eq:omegaNcl}
\omega[N] = \omega^*[N] \, \left\{ \left[ 1 - \eta \frac{f'(\eta)}{f(\eta)} \right]^2 + \left[ v'(n) - f''(\eta) \, \left(\frac{b}{4}\right)^2 \, p^* \right] \frac{n \, \omega^*[N]}{T} \right\}^{-1},
}
where $\omega^*[N] = \chi_2^* / \chi_1^*$ is the scaled variance ofc particle number fluctuations in the ideal gas. For the VDW model ($f(\eta) = 1 - 4\eta$, $u(n) = -a \, n$), the scaled variance is
\eq{
\omega[N] = \omega^*[N] \, \left[ (1-bn)^{-2} - \frac{2 a \,n \, \omega^*[N]}{T} \right]^{-1},
}
in perfect agreement with the previous results of Ref.~\cite{Vovchenko:2015pya}.

The calculation of the higher order cumulants proceeds by iteratively differentiating equations \eqref{eq:c21} and \eqref{eq:c22} with respect to the chemical potential, and by solving the resulting system of equations for the higher-order derivatives of $n$ and $\mu^*$. 

The following equations are obtained for the second-order derivatives $(\partial^2 n / \partial \mu^2)$ and $(\partial^2 \mu^* / \partial \mu^2)$, which are needed to calculate the $\chi_3$:
\begin{alignat}{3}
\left[ 1 - \frac{b}{4} \, f'(\eta) \, n^* \right] &\left(\frac{\partial^2 n}{\partial \mu^2}\right) \quad & - & & f(\eta) \, \chi_2^* \, T^2 \, & \left(\frac{\partial^2 \mu^*}{\partial \mu^2}\right) = b_1^{(2)},  
\label{eq:c31}
\\
\left[ -\left(\frac{b}{4}\right)^2 \, f''(\eta) \, p^* + v'(n) \right] &\left(\frac{\partial^2 n}{\partial \mu^2}\right) \quad & + & & \quad \left[ 1 - \frac{b}{4} \, f'(\eta) \, n^* \right] &\left(\frac{\partial^2 \mu^*}{\partial \mu^2}\right) = b_2^{(2)}.
\label{eq:c32}
\end{alignat}
with
\eq{
b_1^{(2)} & = \left( \frac{b}{4} \right)^2 f''(\eta) \, n^* \left( \frac{\partial n}{\partial \mu} \right)^2
+ 2 \frac{b}{4} f'(\eta) \, \chi_2^* \, T^2 \, \left(\frac{\partial n}{\partial \mu}\right) \left(\frac{\partial \mu^*}{\partial \mu}\right)
+ f(\eta) \, \chi_3^* \, T \, \left( \frac{\partial \mu^*}{\partial \mu} \right)^2, \\
b_2^{(2)} & = \left[ \left( \frac{b}{4} \right)^3 f'''(\eta) \, p^* - v''(n) \right] \left( \frac{\partial n}{\partial \mu} \right)^2
+ 2 \left( \frac{b}{4} \right)^2 f''(\eta) \, n^* \, \left(\frac{\partial n}{\partial \mu}\right) \left(\frac{\partial \mu^*}{\partial \mu}\right)
+ \frac{b}{4}\, f'(\eta) \, \chi_2^* \, T^2 \, \left( \frac{\partial \mu^*}{\partial \mu} \right)^2.
}

Finally, the system of equations for the third-order derivatives $(\partial^3 n / \partial \mu^3)$ and $(\partial^3 \mu^* / \partial \mu^3)$, needed for the calculation of the $\chi_4$, is the following
\begin{alignat}{3}
\left[ 1 - \frac{b}{4} \, f'(\eta) \, n^* \right] &\left(\frac{\partial^3 n}{\partial \mu^3}\right) \quad & - & & f(\eta) \, \chi_2^* \, T^2 \, & \left(\frac{\partial^3 \mu^*}{\partial \mu^3}\right) = b_1^{(3)},  
\label{eq:c41}
\\
\left[ -\left(\frac{b}{4}\right)^2 \, f''(\eta) \, p^* + v'(n) \right] &\left(\frac{\partial^3 n}{\partial \mu^3}\right) \quad & + & & \quad \left[ 1 - \frac{b}{4} \, f'(\eta) \, n^* \right] &\left(\frac{\partial^3 \mu^*}{\partial \mu^3}\right) = b_2^{(3)}.
\label{eq:c42}
\end{alignat}
with
\eq{
b_1^{(3)} & = 
\left( \frac{b}{4} \right)^3 f'''(\eta) \, n^* \left( \frac{\partial n}{\partial \mu} \right)^3
+ 3 \, \left( \frac{b}{4} \right)^2 f''(\eta)
\, \chi_2^* \, T^2 \,  \left(\frac{\partial \mu^*}{\partial \mu}\right) \, \left( \frac{\partial n}{\partial \mu} \right)^2
+
3 \, \left( \frac{b}{4} \right)^2 f''(\eta)
\, n^* \, \left(\frac{\partial^2 n}{\partial \mu^2}\right)  \left( \frac{\partial n}{\partial \mu} \right)
\nonumber \\
& \quad 
+ 
3 \, \frac{b}{4}  f'(\eta)
\,  \chi_2^* \, T^2 \, \left(\frac{\partial^2 \mu^*}{\partial \mu^2} \right) \left( \frac{\partial n}{\partial \mu} \right)
+
3 \, \frac{b}{4} f'(\eta)
\,  \chi_2^* \, T^2 \, \left( \frac{\partial^2 n}{\partial \mu^2} \right) \, \left( \frac{\partial \mu^*}{\partial \mu} \right)
+
3 \, f(\eta)
\,  \chi_3^* \, T \, \left( \frac{\partial^2 \mu^*}{\partial \mu^2}  \right) \left( \frac{\partial \mu^*}{\partial \mu} \right)
\nonumber \\
& \quad +
3 \, \frac{b}{4}  f'(\eta)
\,  \chi_3^* \, T \, \left( \frac{\partial n}{\partial \mu} \right) \left( \frac{\partial \mu^*}{\partial \mu} \right)^2
+ 
f(\eta) \, \chi_4^* \, \left( \frac{\partial \mu^*}{\partial \mu} \right)^3,
}
and
\eq{
b_2^{(3)} & = 
\left[ \left( \frac{b}{4} \right)^4 f''''(\eta) \, p^* - v'''(n) \right] \left( \frac{\partial n}{\partial \mu} \right)^3
+
3 \left( \frac{b}{4} \right)^3
\, f'''(\eta) \,   n^* \left( \frac{\partial \mu^*}{\partial \mu} \right) \left( \frac{\partial n}{\partial \mu} \right)^2
\nonumber \\
& \quad +
\left[3 \, \left( \frac{b}{4} \right)^3 f'''(\eta)
\, p^* - 3 v''(n) \right] \left( \frac{\partial^2 n}{\partial \mu^2} \right) \left( \frac{\partial n}{\partial \mu} \right)
+
3 \left( \frac{b}{4} \right)^2
\, f''(\eta) \,   n^* \left( \frac{\partial^2 \mu^*}{\partial \mu^2} \right) \left( \frac{\partial n}{\partial \mu} \right)
\nonumber \\
& \quad +
3 \left( \frac{b}{4} \right)^2
\, f''(\eta) \,   n^* \left( \frac{\partial^2 n}{\partial \mu^2} \right) \left( \frac{\partial \mu^*}{\partial \mu} \right)
+
3 \frac{b}{4}
\, f'(\eta) \,  \chi_2^* \, T^2 \, \left( \frac{\partial^2 \mu^*}{\partial \mu^2} \right) \left( \frac{\partial \mu^*}{\partial \mu} \right)
\nonumber \\
& \quad +
3 \left( \frac{b}{4} \right)^2
\, f''(\eta) \,   \chi_2^* \, T^2 \, \left( \frac{\partial n}{\partial \mu} \right) \left( \frac{\partial \mu^*}{\partial \mu} \right)^2
+ 
\frac{b}{4} \, f'(\eta) \,  \chi_3^* \, T \, \left( \frac{\partial \mu^*}{\partial \mu} \right)^3.
}

\end{widetext}

\end{document}